\documentclass[%
 reprint,
groupedaddress,
nofootinbib,
 amsmath,amssymb,
 aps,
]{revtex4-1}
\usepackage{scrextend}

\usepackage{soul}
\usepackage{graphicx}
\usepackage{dcolumn}
\usepackage{bm}
\usepackage{hyperref}
\usepackage{xcolor}

\usepackage[normalem]{ulem}

\begin{document}

\preprint{APS/123-QED}

\title{Parameterizing Noise Covariance in Maximum-Likelihood Component Separation}

\author{Goureesankar Sathyanathan}
 \email{gouree.sath@gmail.com}
\affiliation{ Indian Institute of Science Education and Research Thiruvananthapuram, 69551, Kerala, India}
\affiliation{Department of Physics and Astronomy, University of California, Riverside, 92521, California, USA
}

\author{Josquin Errard}
\affiliation{Universit\'e Paris Cit\'e, CNRS, Astroparticule et Cosmologie, F-75013 Paris, France}

\author{Soumen Basak}
\affiliation{Indian Institute of Science Education and Research Thiruvananthapuram, 69551, Kerala, India
}%

\date{\today}

\begin{abstract}
We introduce a noise-aware extension to the parametric maximum-likelihood framework for component separation by modeling correlated \(1/f^\alpha\) noise as a harmonic-space power law. This approach addresses a key limitation of existing implementations, for which a mismodelling of the statistical properties of the noise can lead to biases in the characterization of the spectral laws, and consequently biases in the recovered CMB maps.  We propose a novel framework based on a modified ridge likelihood embedded in an ensemble-average pipeline and derive an analytic bias correction to control noise-induced foreground residuals. We discuss the practical applications of this approach in the absence of true noise information, leading to the choice of white noise as a realistic assumption. As a proof of concept, we apply this methodology to a set of simplified, idealized simulations inspired by the specifications of the proposed ECHO (CMB-Bh\(\overline{a}\)rat) mission, which features multi-frequency, large-format focal planes. We forecast the $95 \%$ upper limit on the tensor-to-scalar ratio, \(r_{95}\), under a suite of realistic noise scenarios.  Our results show that for an optimistic full sky observation, ECHO can achieve \(r_{95}\leq 10^{-4}\) even in the presence of significant correlated noise, demonstrating the mission's capability to probe primordial gravitational waves with unprecedented sensitivity. Without degrading the statistical performance of the traditional component separation, this methodology offers a robust path toward next-generation B-mode searches and informs instrument design by quantifying the impact of noise correlations on cosmological parameter recovery.
\end{abstract}

\maketitle

\section{Introduction}

The Cosmic Microwave Background (CMB) opens a window into the earliest moments of the universe and guides our understanding of its subsequent development. Its serendipitous discovery by Penzias and Wilson in the mid-1960s \cite{1965ApJ...142..419P} transformed observational cosmology. Since then, a succession of ground- and space-based experiments has mapped the tiny temperature variations and polarization of the CMB over a wide range of angular scales and frequencies \cite{2011A&A...536A...1P, P_A_R_Ade_2014, 2020JCAP...12..047A, 2020JCAP...12..045C, Sayre_2020, 2018JLTP..193.1112G, 2020_2}. These measurements established the standard Lambda Cold Dark Matter concordance model ($\Lambda$CDM)\cite{2008cosm.book.....W, Bull_2016} and have since provided rigorous tests of models for dark matter, dark energy, and the very fabric of space-time~\cite{Peebles2009Finding, 2014PTEP.2014fB101S, 2020PTEP.2020h3C01P}.

As the measurement of the temperature anisotropies has reached its practical limits \cite{refId0,Komatsu_2014}, attention has turned to the much fainter polarization signals. These polarization measurements provide a complementary window into the early universe, carrying additional information about the primordial physics that temperature data alone cannot reveal. Both the temperature and polarization patterns originate from tiny perturbations of the space-time metric produced during a brief episode of accelerated expansion, or cosmic inflation \cite{1979JETPL..30..682S,1980ApJ...241L..59K,1981PhRvD..23..347G,1981MNRAS.195..467S,1982PhLB..108..389L,1982PhRvL..48.1220A}. In that hypothetical phase, quantum fluctuations were stretched from microscopic to cosmological scales, seeding both the density variations that led to structure formation and a background of primordial gravitational waves. The latter imprints a characteristic parity-odd curl-type polarization pattern, known as B-modes, on the CMB sky \cite{Kamionkowski_1997, Seljak_1997,Hu_1997}.

In recent years, CMB research has turned to the search for primordial B-modes, marking one of the next major challenges in cosmology. These B-modes could offer a cleaner window into the earliest stages of our universe, probing energy scales far beyond the reach of laboratory experiments on the ground. Although the exact amplitude of the signal remains uncertain, theory robustly predicts a two-peak structure in the B-mode power spectrum. The first peak, at very large angular scales ($\ell < 10$), arises from scattering during the reionisation epoch. The second, at degree scales, reflects processes at matter-radiation decoupling during recombination.

The amplitude of primordial B-modes is quantified by the tensor-to-scalar ratio \(r\), which different inflationary scenarios predict to be \(r\lesssim10^{-1}\) \cite{2013arXiv1303.3787M, Martin_2024}.  To date, no unambiguous detection has been made. The pursuit of a precise measurement of the tensor-to-scalar ratio \(r\) has driven a coordinated global effort, spanning both ground-based observatories and spaceborne experiments, and has steadily lowered the upper bound. The Planck 2018 analysis set \(r_{0.002}<0.10\) (95 \% CL) using temperature, polarization, and lensing data~\cite{planck2018}. Combining Planck with BICEP2/Keck BK15 tightened this to \(r_{0.05}<0.062\)~\cite{bicepkeck2018}, and the latest BK18 joint analysis with Planck PR4 gives \(r_{0.05}<0.032\) (95 \% CL) at the pivot \(k=0.05\,\mathrm{Mpc}^{-1}\)~\cite{Tristram_2022}.  

Ground-based results from SPTpol, SPIDER, ACTPol and POLARBEAR provide complementary bounds $r < \mathcal{O}(10^{-2})$ \cite{sptpol2018,polarbear2017, 2022ApJ...927..174A, louis2025atacamacosmologytelescopedr6}, and the next generation experiments like the Simons Observatory and CMB-S4 aim to reach \(r\leq \mathcal{O}(10^{-3})\) sensitivities~\cite{simons2019,cmbs42016}. Complementing these efforts, the LiteBIRD satellite aims to achieve high-fidelity measurements of large-scale \(B\)-modes with stringent control of systematics~\cite{2019JLTP..194..443H}.

Building on these achievements and the heritage of the COrE concept~\cite{thecorecollaboration2011corecosmicoriginsexplorer}, the Exploring Cosmic History and Origins (ECHO) mission, also known as CMB-Bh\(\overline{a}\)rat, currently under active review at the Indian Space Research Organisation (ISRO)\footnote{\url{https://www.isro.gov.in/}}, seeks to extend these capabilities. It is designed to transform upper limits into a robust detection or a \(\sigma(r) < 10^{-3}\) constraint on $r$.  The instrument will carry multi-frequency, large-format focal-plane arrays ($20-800$ GHz) on a satellite platform comprising thousands of detectors, employing rapid scan strategies to suppress low-$\ell$ noise. Forecasts indicate \(\sigma(r)\approx10^{-4}\) even in the absence of a detection, allowing ECHO to decisively test a broad class of inflationary models and probe the gravitational-wave background with unprecedented precision and sensitivity, representing a major milestone in our ability to probe the energy scale of inflation~\cite{Adak_2022,sen2023importancehighfrequencybandsthermal}.

The primordial B-mode signal is expected to be extremely faint, at least an order of magnitude below current upper bounds. At large angular scales, true B-modes arise solely from tensor perturbations associated with primordial gravitational waves. However, several confounding effects complicate their detection (see~\cite{Errard_2016} for details). Chief among these is gravitational lensing, where the deflection of CMB photons by intervening large-scale structures remaps the primary anisotropies~\cite{LEWIS_2006}. This remapping induces non-negligible distortions on sub-degree scales and transfers a fraction of the dominant parity-even E-mode polarization patterns into B-mode, boosting the small-scale power of the latter and thereby masking the primordial B-mode signal~\cite{PhysRevD.58.023003, 2002ApJ...574..566H}. Consequently, delensing, the process of removing the lensing-induced B-modes (see \cite{2017JCAP...05..035C} for details) has become a central element in the strategy of future satellite missions such as ECHO.

In addition to primordial B-mode signals being intrinsically faint, the polarized emission from our own Galaxy is several orders of magnitude stronger and poses a major challenge to their detection~\cite{Krachmalnicoff_2016, Errard_2016}. The principal sources of this foreground contamination are synchrotron radiation, which dominates at low frequencies, $\nu\, \leq\, 70$ GHz, and thermal dust emission, which becomes significant at higher frequencies, $\nu\, \geq\, 100$ GHz. Both mechanisms exhibit distinct spectral energy distributions (SEDs), allowing for their separation from the CMB via multi-frequency observations. Parametric component separation techniques attempt to exploit these spectral differences by modeling each component's SED and fitting the parameters using observational data~\cite{cardoso2008componentseparationflexiblemodels,Stompor_2009}. However, imperfect modeling due to, e.g. spatial variation in foreground properties, leads to residual contamination in the reconstructed CMB maps. These residuals are particularly critical when targeting constraints on the tensor-to-scalar ratio at the level of \(r \sim 10^{-3}\), where even small mismodeling of the foregrounds or instrumental response can introduce biases of multiple standard deviations in the inferred value of \(r\)~\cite{Betoule:2009pq,Errard_2012,Errard_2016,Stompor_2016,Errard_2019,thorne2019removalgalacticforegroundssimons,2022ApJ...926...54A,2023PTEP.2023d2F01L}. Consequently, robust mitigation of foreground-induced systematics is a necessary precondition for achieving reliable measurements of primordial B-modes in next-generation experiments.

Applying component separation techniques to real observational data requires a comprehensive treatment of instrumental noise and associated systematics~\cite{Natoli_2018,Abitbol_2021}. While thermal detector noise is well-characterized by Johnson-Nyquist theory and is typically modeled as a Gaussian process uniform across frequency channels and angular scales \cite{PhysRev.32.110,PhysRev.32.97}, actual observations reveal a more complex noise landscape. In practice, additional noise contributions arise from low-frequency drifts in the readout electronics, variations in instrument gain, and atmospheric fluctuations or ground-pickup, particularly relevant for ground-based observatories. These effects induce temporal and spatial correlations that are not captured by idealized white-noise models. Neglecting such correlations can lead to significant biases in both the reconstruction of the CMB signal and the estimation of foreground parameters ~\cite{Stompor_2009, Errard_2012}. Accurate modeling of the noise covariance structure is therefore critical for reliable component separation, especially in the low-noise regime relevant for next-generation CMB experiments targeting primordial B-modes.\\

In this work, we introduce a harmonic-space power-law model to describe the noise covariance, motivated by the stationarity and correlated noise characteristics observed in modern CMB experiments. We extend the standard spectral likelihood framework by developing a modified ridge likelihood that allows us to adjust the noise parameters along with the foreground ones, and incorporates an analytic bias correction to account for noise-induced distortions. We discuss the practical limitations of this approach due to the unavailability of true noise characteristics and propose a reasonable white noise assumption, in line with the strategy used in current CMB missions \cite{2023PTEP.2023d2F01L, 2024A&A...686A..16W}. To facilitate efficient forecasting, we construct an ensemble-average likelihood pipeline that avoids reliance on Monte Carlo realizations. Using this pipeline, we estimate the projected $95\%$ upper limit on the tensor-to-scalar ratio, \(r_{95}\), achievable with the specifications of the upcoming ECHO mission.

The structure of the paper is as follows. Section~\ref{sec:methodology} introduces the noise-aware extension to the maximum-likelihood parametric component separation framework, including the formulation of a modified likelihood. We also present an analytical bias correction and discuss strategies for practical implementation. Section~\ref{sec:pipeline} describes the ensemble-average implementation and outlines the steps involved in the parameter recovery. Forecasted constraints on the tensor-to-scalar ratio \(r\), derived using this pipeline and representative instrument specifications, are presented in Section~\ref{sec:results}. We finally summarize and discuss our findings in Section~\ref{sec:conclusions}. 

\section{Methodology} \label{sec:methodology}
Based upon the concepts introduced earlier, we now present the formal framework for our noise-aware component separation exercise. We begin by reviewing the standard spectral likelihood and then introduce a harmonic-space, power-law model for correlated noise. This noise model is incorporated into a modified ridge likelihood, which we implement within an ensemble-average pipeline combined with an analytic bias correction. Together, these elements form the basis for our forecasts of the tensor-to-scalar ratio \(r\) under the ECHO mission specifications.

\subsection{Spectral likelihood}

We model the multi-frequency data as a linear combination of sky components and instrument noise:
\begin{equation}
\label{eq:basic_eqn}
    \mathbf{d} = \mathbf{A}\,\mathbf{s} + \mathbf{n} .
\end{equation}
Here, \(\mathbf{d}\) is the concatenated data vector containing all \(n_s\) Stokes parameters across \(n_f\) frequency channels.  The vector \(\mathbf{s}\) holds the true sky signals for each of the \(n_c\) sky components, the term \(\mathbf{n}\) represents the instrumental noise affecting each \(n_s\) Stokes parameter across the \(n_f\) channels, and \(\mathbf{A}\) is the \( (n_f n_s)\times(n_c n_s)\) mixing matrix that describes how each component contributes to each frequency.  

Parametric approaches assume specific spectral forms for each sky component, encoding this information in the columns of the mixing matrix \(\mathbf{A}\)~\cite{Eriksen_2006,Stompor_2009,Hoz_2020}.  Introducing a set of spectral parameters \(\beta\), the matrix can be expressed as \(\mathbf{A}\equiv\mathbf{A}(\beta)\).  The task of component separation then reduces to first estimating \(\beta\) from the data, as well as estimating \(\mathbf{s}\), and in particular its CMB component.  
In practice, as detailed below, and following~\cite{Stompor_2009}, the first step can be done by maximizing the spectral likelihood of the observed maps with respect to \(\beta\). The second step, considering the estimated spectral indices and consequently the estimated mixing matrix, is typically performed using a generalized least-square equation. 

The formulation in Equation~\eqref{eq:basic_eqn} is flexible enough to accommodate different mathematical bases, whether real-space maps, spherical-harmonic coefficients, or needlet representations, each offering distinct advantages for specific tasks.  In this work, we adopt the spherical-harmonic domain to streamline the handling of isotropic, correlated noise and beam effects; our motivation for this choice is described in the following sections.

Assuming the instrumental noise is a zero-mean Gaussian with (possibly non-diagonal) covariance matrix \(\mathbf{N}\), the data log-likelihood (up to an additive constant) can be written as
\begin{eqnarray}
\begin{split}
    -2\ln\mathcal{L}_{\rm data}(\mathbf{s},\beta)
&= (\mathbf{d}-\mathbf{A}\,\mathbf{s})^{T}\,\mathbf{N}^{-1}\,\\
&\quad \times(\mathbf{d}-\mathbf{A}\,\mathbf{s}) + \mathrm{const}\,.
\end{split}
\label{eq:full_data_likelihood}
\end{eqnarray}
Equation~\ref{eq:full_data_likelihood} depends on the sky signal \(\mathbf{s}\) and the spectral parameters \(\beta\). Following the two-step procedure of~\citet{Stompor_2009}, we first profile over \(\mathbf{s}\) to obtain the spectral likelihood, then reconstruct the component maps using the best-fit \(\beta\).

Maximizing the data likelihood with respect to \(\mathbf{s}\) leads to the definition of the spectral likelihood
\begin{equation}
\label{eq:spec_likelihood}
\begin{split}
-2\ln\mathcal{L}_{\rm sp}(\beta)
&\equiv-\bigl[\mathbf{A}^{T}\mathbf{N}^{-1}\mathbf{d}\bigr]^{T}
\bigl[\mathbf{A}^{T}\mathbf{N}^{-1}\mathbf{A}\bigr]^{-1}\\
&\quad\times\bigl[\mathbf{A}^{T}\mathbf{N}^{-1}\mathbf{d}\bigr]
+\mathrm{const}\,.
\end{split}
\end{equation}
Once \(\beta\) is fixed at the maximum of the spectral likelihood, we derive the component maps by solving the normal equation of the likelihood, obtained by differentiating \(\ln\mathcal{L}_{\rm data}(\mathbf{s},\beta)\) with respect to \(\mathbf{s}\) giving
\begin{equation}
\mathbf{A}^{T}\,\mathbf{N}^{-1}\,\bigl[\mathbf{d}-\mathbf{A}\,\mathbf{s}\bigr]\;=\;0,
\end{equation}
which leads to the linear system
\begin{equation}
\mathbf{A}^{T}\,\mathbf{N}^{-1}\,\mathbf{A}\,\mathbf{s}
\;=\;\mathbf{A}^{T}\,\mathbf{N}^{-1}\,\mathbf{d}.
\end{equation}
Solving for \(\mathbf{s}\) yields the generalized least-squares estimator:
\begin{equation}
\label{eq:map_recovery}
\widehat{\mathbf{s}}
=\mathbf{W}\,\mathbf{d}\,
\end{equation}
where,
\begin{equation}
    \label{eq:weights}
    \mathbf{W} = \bigl[\mathbf{A}^{T}\,\mathbf{N}^{-1}\,\mathbf{A}\bigr]^{-1} \;\mathbf{A}^{T}\,\mathbf{N}^{-1}\,.\\
\end{equation}
This two-step procedure cleanly separates the estimation of spectral parameters from the reconstruction of the component maps. In the next section, we extend this framework by incorporating a harmonic-space noise model and modifying the ridge likelihood in Equation~\ref{eq:spec_likelihood}, to capture correlated \(1/f^\alpha\) noise in typical CMB datasets.

\subsection{Accounting for noise}  
\label{sec:noise}

Accurate noise modeling and control can be as critical as the treatment of foregrounds in CMB analyses.  Low-frequency drifts in amplifiers, readout electronics, and (for ground-based instruments) atmospheric fluctuations as well as ground pickup introduce correlated noise features that a white-noise assumption cannot capture. In the following, we review the origin of these effects and then translate them into a harmonic-space form suitable for our likelihood framework.

\subsubsection{Modified noise model}

Early work by Schottky \cite{1918AnP...362..541S,PhysRev.28.1331} and Johnson \cite{PhysRev.26.71} demonstrated that resistive components exhibit excess \(1/f^{\alpha}\) noise at low frequencies.  This detector noise power spectral density (PSD) is typically modeled in the frequency domain as
\[
P(f)=P_{\rm white}\,\Bigl[1 + \Big(\tfrac{\displaystyle f_{k}}{\displaystyle f}\Big)^{\alpha}\Bigr],
\]
where \(P_{\rm white}\) is the Johnson-Nyquist term, \(f_{k}\) is the knee frequency, and \(\alpha\) sets the slope of the low-frequency excess.
Large angular scales on the sky correspond naively to low temporal frequencies during scanning, suggesting the empirical mapping
\[
\frac{1}{f^{\alpha}}\;\longrightarrow\;\frac{1}{\ell^{\alpha'}}\,,
\]
where $\alpha'$ and the multiplicative factors would strongly depend on the exact scanning strategy. 
Accordingly, we adopt the following harmonic-space noise angular power spectrum:
\begin{equation}
\label{eq:noise_aps}
\mathbf{N_{\ell}} 
\equiv \mathbf{N_\ell^{(\nu)}}(p)\,
=\sigma_{\rm white}^{2}(\nu)\,
\Bigl[1 + \Big(\tfrac{\displaystyle \ell}{\displaystyle \ell_{0}(\nu)}\Big)^{\alpha(\nu)}\Bigr]\\
.
\end{equation}
Here:\\
\begin{itemize}
    \item \(\sigma_{\rm white}^{2}(\nu)\) is the white-noise variance in frequency channel \(\nu\). 
    \item \(\ell_{0}(\nu)\) is the multipole at which correlated noise matches the white-noise level.  
    \item \(\alpha(\nu)\) controls the steepness of the power law. 
\end{itemize}
Each of the \(n_{f}\) frequency channels may have its own set of parameters $p\equiv\{\sigma_{\rm white}, \ell_{0}, \alpha\}$. While, in practice, \(\sigma_{\rm white}\) can also be chosen as a free parameter, in this work we consider it to be fixed at the noise RMS levels proposed in the ECHO mission specifications. As a result, we consider $p\,=\,\{\ell_{0},\alpha\}$ as the free parameters describing the noise angular power spectrum.

\begin{figure}
  \centering
  \includegraphics[width=\columnwidth]{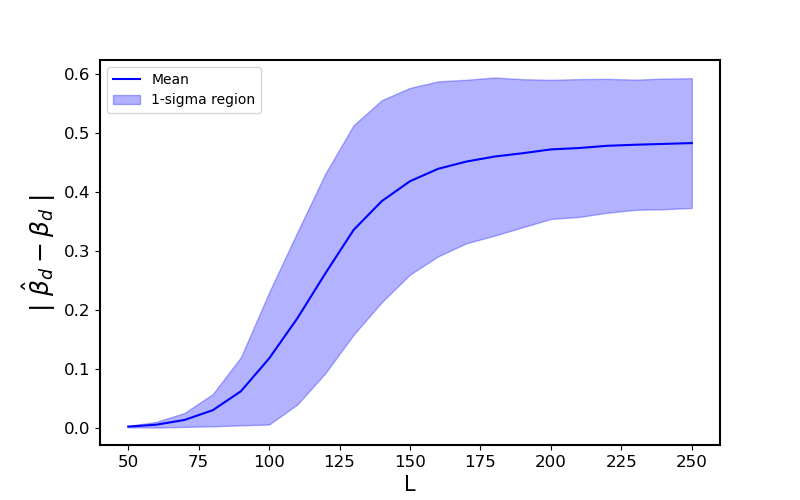}
  \caption{Bias in the recovered dust spectral index \(\hat\beta_{d}\) when correlated noise is ignored, considering only white noise in the evaluation of the spectral likelihood, Equation~\ref{eq:spec_likelihood}.  The input noise follows Equation~\eqref{eq:noise_aps} with \(\alpha=-6\), and \(\ell_{0}\) is drawn \(100\times n_{f}\) times from a uniform interval \([\ell_{\min}=2,\ell_{\max}=L]\).  As \(L\) increases, the noise model's complexity grows and the bias on \(\hat\beta_{d}\) increases when using a white-noise likelihood. The blue curve represents the average bias over the 100 simulations.}
  \label{fig:incorrect_estimate}
\end{figure}

Figure~\ref{fig:incorrect_estimate} illustrates how neglecting correlated noise in the evaluation of the spectral likelihood, Equation~\ref{eq:spec_likelihood}, leads to biased estimates of the dust index \(\beta_{d}\).  In this example, we draw \(\ell_{0}\) values for each frequency band from a \([\ell_{\min}=2,\ell_{\max}=L]\) range, with increasing $L$ values while keeping \(\alpha=-6\) to amplify the effect.  Although these changes affect the noise of the observations, we force the spectral likelihood to use a white noise covariance -- diagonal $\mathbf{N}$ in real space. As the possible range for \(\ell_{0}\) increases, the noise complexity of the data increases, and the standard spectral likelihood produces increasingly biased \(\hat\beta_{d}\).  These biases in the foreground parameters turn eventually into residuals that can distort the tensor-to-scalar ratio \(r\), particularly on the large scales targeted by ECHO. We note that only a frequency-varying mismatch between the true and assumed noise covariance can, in this context, lead to a biased characterization of the spectral parameters. This remark will remain true for the rest of the paper.\\

To mitigate such biases, we embed the power-law noise model of Equation~\eqref{eq:noise_aps} into a modified ridge likelihood and derive an analytic bias correction in Section~\ref{sec:ridge_likelihood}.  This extension should ensure that correlated noise is properly taken into account in parameter estimation, mitigating the biases shown in Figure~\ref{fig:incorrect_estimate}.

\subsubsection{Modified ridge likelihood} \label{sec:ridge_likelihood}

In the spherical-harmonic domain, the data log-likelihood reads
\begin{equation}
\begin{split}
\label{eq:original_likelihood}
-2\ln\mathcal{L}_{\rm data}(s_{\ell m},\beta,p)
&=\sum_{\ell m}\bigl[\mathbf{d}_{\ell m}-\mathbf{A}\,\mathbf{s}_{\ell m}\bigr]^{\!\dagger}\
\mathbf{N}_{\ell}^{-1}\,\\
&\quad \times\bigl[\mathbf{d}_{\ell m}-\mathbf{A}\,\mathbf{s}_{\ell m}\bigr]\\
&\quad+\sum_{\ell}(2\ell+1)\,\ln\bigl|\mathbf{N}_{\ell}\bigr|
+\mathrm{const}\,.
\end{split}
\end{equation}
Profiling over the component amplitudes \(\mathbf{s}_{\ell m}\) yields the ridge likelihood, which now depends only on the parameters \(\beta\) and \(p\):
\begin{equation}
\label{eq:RidgeLikelihood}
\begin{split}
-2\ln\mathcal{L}_{\rm ridge}(\beta,p)
&=-\sum_{\ell m}\bigl[\mathbf{A}^{T}\mathbf{N}_{\ell}^{-1}\mathbf{d}_{\ell m}\bigr]^{\!\dagger}
\bigl[\mathbf{A}^{T}\mathbf{N}_{\ell}^{-1}\mathbf{A}\bigr]^{-1}\\
&\quad\times\bigl[\mathbf{A}^{T}\mathbf{N}_{\ell}^{-1}\mathbf{d}_{\ell m}\bigr]
+\sum_{\ell m}\mathbf{d}_{\ell m}^{\dagger}\mathbf{N}_{\ell}^{-1}\mathbf{d}_{\ell m}\\
&\quad+\sum_{\ell}(2\ell+1)\,\ln\bigl|\mathbf{N}_{\ell}\bigr|
+\mathrm{const}\,.
\end{split}
\end{equation}
Maximizing \(\mathcal{L}_{\rm ridge}\) over these parameters provides the global likelihood peak \((\hat\beta,\hat p)\) without fitting \(\mathbf{s}_{\ell m}\) explicitly.
\begin{equation}
\label{eq:bestfit}
(\hat\beta,\hat p) \;=\;{\rm argmin}_{\beta,p}\bigl[-2\ln\mathcal{L}_{\rm ridge}(\beta,p)\bigr].
\end{equation}

\begin{figure}
  \centering
  \includegraphics[width=\columnwidth]{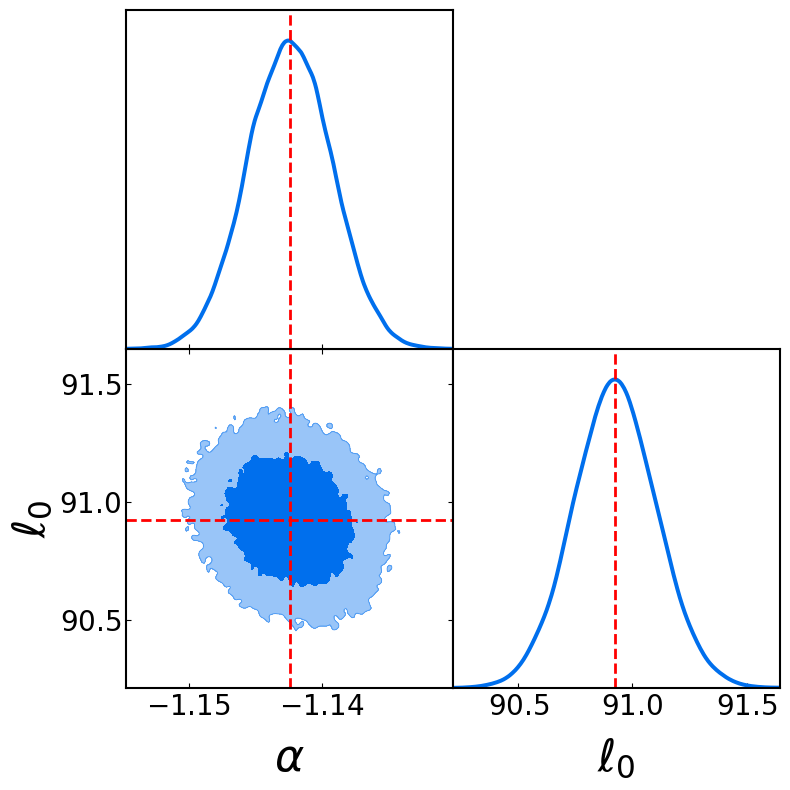}
  \caption{Joint posterior contours (68 \% and 95 \% confidence levels) for the noise parameters \(\alpha\) and \(\ell_{0}\) marginalized over the foreground spectral parameters (\(\beta_{d}\), \(T_{d}\), \(\beta_{s}\)) obtained by sampling the ensemble averaged version of the modified ridge likelihood (Equation~\ref{eq:RidgeLikelihood}).  Dashed lines mark the recovered values --- \( \alpha \) and \( \ell_0\) have bias of \(\sim 50 \sigma\) and \( \sim 200 \sigma\) respectively, when compared to their true values of $\alpha = -1, \;\ell_0 = \ell_{\max}/2 = 128$.}
  \label{fig:contour}
\end{figure}

This yields the best-fit spectral and noise parameters without explicitly fitting the component maps, as for the original spectral likelihood.
Figure~\ref{fig:contour} shows the ridge-likelihood contours obtained from Equation~\eqref{eq:RidgeLikelihood} for a representative case (\(\alpha=-1\), \(\ell_{0}=128\)), marginalized over the spectral parameters \( \beta \). The peak of the likelihood (red broken lines) for these noise parameters is offset from the true input parameters roughly by \(\sim 50\sigma\) for \(\alpha\) and  \(\sim200\sigma\) for \(\ell_0\), indicating an important, systematic bias in the procedure.
This bias arises because, when the map estimate \(\widehat{\mathbf{s}}_{\ell m}\) is substituted back into the full data likelihood, the effective degrees of freedom are reduced, leading to a misestimate of the noise covariance. However, it is known that for the unbiased recovery of the spectral parameters \(\beta_{\rm true}\), the weights should match exactly the mixing matrix such that \(\mathbf{WA = I}\). Inserting the expression of \(\widehat{\mathbf{s}}_{\ell m}\) from Equation~\eqref{eq:map_recovery} into the original likelihood, we get
\begin{equation}
\label{eq:missing_temrs}
\begin{split}
-2\ln\mathcal{L}_{\rm data}(\beta_{\rm true},p)
&=\sum_{\ell m}\bigl[\mathbf{d}_{\ell m}-\mathbf{A}\,\widehat{\mathbf{s}}_{\ell m}\bigr]^{\!\dagger}
\mathbf{N}_{\ell}^{-1}\\
&\quad\times\bigl[\mathbf{d}_{\ell m}-\mathbf{A}\,\widehat{\mathbf{s}}_{\ell m}\bigr]\\
&\quad +\sum_{\ell}(2\ell+1)\ln\lvert\mathbf{N}_{\ell}\rvert+\mathrm{const}\\
&=\sum_{\ell m}\bigl[(\mathbf{I-P})\,\mathbf{n}_{\ell m}\bigr]^{\!\dagger}
\mathbf{N}_{\ell}^{-1}\,\\
&\quad \times \bigl[(\mathbf{I-P})\,\mathbf{n}_{\ell m}\bigr]\\
&\quad+\sum_{\ell}(2\ell+1)\ln\lvert\mathbf{N}_{\ell}\rvert+\mathrm{const}.
\end{split}
\end{equation}
Here, $\mathbf{I}$ is the identity matrix and we define the projection operator \(\mathbf{P} \equiv \mathbf{AW}\) so that the matrix \(\mathbf{I-P}\) removes the modes estimated in the profiling step, effectively reducing the amplitude of the noise level. This wrong estimation of the noise level leads to inaccurate estimates of the spectral parameters, as shown in Figure \ref{fig:incorrect_estimate}, and this manifests as significant residuals at the power spectrum level.

Since the analytical expression of the bias is known, we can correct for it in the ridge likelihood by incorporating the information that is lost when \({\mathbf{s}}_{\ell m}\) is replaced by \(\widehat{\mathbf{s}}_{\ell m}\) during profiling. The correction accounts for the reduction in noise covariance produced by the projection of the noise onto the space spanned by \(\mathbf{P}\).

It is straightforward to show that the operator \(\mathbf{P}\) satisfies the following identities:
\begin{itemize}
    \item Idempotent, with \(\mathbf{P^2 = P}\),
    \item Self-adjoint, with \(\mathbf{P}^T\; \mathbf{N}_{\ell}^{-1} = \mathbf{N}_{\ell}^{-1} \; \mathbf{P}\).
\end{itemize} 
Using these identities, the bias-corrected ridge likelihood becomes  

\begin{eqnarray}
\label{eq:bias_corrected_ridge_likelihood}
\begin{split}
&-2\ln\mathcal{L}_{\rm corrected}(\beta,p)
=-2\ln\mathcal{L}_{\rm ridge}(\beta,p)\\
&+\sum_{\ell m}\Bigl[\mathbf{n}_{\ell m}^{\dagger} \mathbf{P}^{T}\mathbf{N_{\ell}^{-1}n}_{\ell m}\\
&\;+\,\mathbf{n_{\ell m}^{\dagger}N_{\ell}^{-1}P\,n}_{\ell m}\Bigr]\\
&-\sum_{\ell m}\mathbf{n}_{\ell m}^{\dagger}\mathbf{P}^{T}\mathbf{N_{\ell}^{-1}P\,n}_{\ell m}\,.\\
&=-2\ln\mathcal{L}_{\rm ridge}(\beta,p) +\sum_{\ell m}\mathbf{n}_{\ell m}^{\dagger} \mathbf{N}_{\ell}^{-1} \mathbf{P} \,\mathbf{n}_{\ell m}
\end{split}
\end{eqnarray} 

These correction terms reintroduce the noise modes that were suppressed and restore an unbiased estimate of the noise covariance. It is important to note that these are additional terms arising out of Equation \eqref{eq:missing_temrs} and do not cancel out any terms in the original ridge likelihood, Equation \eqref{eq:RidgeLikelihood}.

Practical implementation of the new estimator requires careful consideration of several factors. First, in all realistic scenarios, the true noise \(\mathbf{n_{\ell m}}\) is not directly accessible. If the true noise indeed follows the form modeled in Equation \eqref{eq:noise_aps}, then determining it would require knowledge of the noise parameters \(p\) that characterize its behavior, parameters which are not known a priori. This can be addressed by considering that the theoretical noise terms in the bias-corrected ridge likelihood are constructed using a simple white noise. Moreover, the quantities appearing in Equation \eqref{eq:bias_corrected_ridge_likelihood}, namely, the spherical harmonic coefficients, are not always accessible in practice. Evaluating them would require the use of a large number of simulations, making the approach computationally expensive. Alternatively, we semi-analytically estimate the ensemble average limit to replace the spherical harmonics coefficients with their power spectrum, as discussed below.

\section{Pipeline}  
\label{sec:pipeline}

Having established a noise-covariance-aware, bias-corrected ridge likelihood, we now describe its implementation without the need for Monte Carlo realizations. In the ensemble-average limit, corresponding to an infinite number of signal+noise simulations, and similarly to what was done in~\cite{Stompor_2016}, we replace the sums over spherical-harmonic modes with traces over theoretical covariance matrices, and construct a likelihood that retains full statistical fidelity for the CMB, foregrounds, and instrument noise while drastically reducing computational requirements.

\subsection{Ensemble-average likelihood}

Starting from the bias-corrected ridge likelihood, where each term is a scalar and can be written as a trace, the ensemble-averaged version can be obtained as:

\begin{equation}
\label{eq:final_eq}
\begin{split}
\bigl\langle -2\ln\mathcal{L}_{\rm corrected}\bigr\rangle
&=\sum_{\ell}(2\ell+1)\mathrm{Tr}\bigl[\mathbf{N}_{\ell}^{-1}\,
\mathbf{D}_\ell \bigr]\\
&\quad-\sum_{\ell}(2\ell+1)\mathrm{Tr}\bigl[\mathbf{N}_{\ell}^{-1}\,
\mathbf{A}\,(\mathbf{A}^{T}\mathbf{N}_{\ell}^{-1}\mathbf{A})^{-1}\\
&\qquad\quad\times\mathbf{A}^{T}\mathbf{N}_{\ell}^{-1}\,
\mathbf{D}_\ell \bigr]\\
&\quad+\sum_{\ell}(2\ell+1)\,\ln\bigl|\mathbf{N}_{\ell}\bigr| \\
&\quad +\sum_{\ell }(2\ell+1)\mathrm{Tr}\bigl[\mathbf{N}_{\ell}^{-1}\ \mathbf{P} \
 \mathbf{N}_{\ell}^{\rm th}\bigr]\\
\end{split}
\end{equation}

The theoretical ensemble average estimator for a given pair of frequencies is given as

\begin{align}
\mathbf{D}_\ell =  \frac{1}{2 \ell + 1}\bigl\langle \sum_{m}\mathbf{d}_{\ell m}\mathbf{d}_{\ell m}^{\dagger}\bigr\rangle
&=\mathbf{C}_{\ell}^{\rm CMB}+\mathbf{C}_{\ell}^{\rm FG}+\mathbf{N}_{\ell}
\end{align}
where \(\mathbf{C}_{\ell}^{\rm CMB}\) and \(\mathbf{C}_{\ell}^{\rm FG}\) are the theoretical power spectra, obtained from the Code for Anisotropies in the Microwave Background (\textsc{CAMB}) \cite{2011ascl.soft02026L} and Python Sky Model (\textsc{PySM}) \cite{10.1093/mnras/stx949} respectively. The input noise power spectrum \(\mathbf{N}_{\ell}\) is obtained using the model specified by Equation~\eqref{eq:noise_aps} and represent different possible noise configurations as mentioned in Section \ref{sec:results}.

If we had access to the true input noise, which we assume to be of the form given by Equation \eqref{eq:noise_aps}, then we could have used \begin{eqnarray}\label{eq:pink_noise_th}
    \mathbf{N}_{\ell}^{\rm th} = \frac{1}{2 \ell + 1}\bigl\langle\sum_{m}\mathbf{n}_{\ell m}\mathbf{n}_{\ell m}^{\dagger}\bigr\rangle= \mathbf{N_\ell^{(\nu)}}(p).
\end{eqnarray}
This would require the knowledge of the power-law index \(\alpha\) and the knee multipole \( \ell_0\) of the true noise present in the data. However, since we never have access to the theoretical noise in practice, it is not possible to move forward under this assumption. Hence, we choose to adopt, as the theoretical noise model, a simple white noise configuration given by the specifications mentioned in the ECHO proposal as \begin{eqnarray}
\label{white_noise_th}
    \mathbf{N}_{\ell}^{\rm th} = \frac{1}{2 \ell + 1}\bigl\langle\sum_{m}\mathbf{n}_{\ell m}\mathbf{n}_{\ell m}^{\dagger}\bigr\rangle
= \sigma^2_{\rm white}(\nu).
\end{eqnarray}
Although we understand that this is one of the simplest assumptions we can make about the characteristics of the noise present in the data, this represents a technological limitation at present. Recent works by the LiteBIRD collaboration \cite{2023PTEP.2023d2F01L} and the Simons Observatory \cite{2024A&A...686A..16W} employ white noise in their analysis as well. This allows us to study the extent to which the choice of bias correction and the subsequent theoretical noise model influences the recovery of the noise and spectral parameters.

We restrict our study to the spatially invariant foreground model (\texttt{d0s0}), leaving extensions to spatially varying SEDs to future work. Sampling from the bias-corrected ridge likelihood over the parameter space \(\{\beta,p\}\), using the Metropolis-Hastings algorithm from \textsc{emcee} \cite{2013ascl.soft03002F}, then yields the best-fit spectral and noise parameters without the need for any simulations.
Using the best-fit parameters \((\hat\beta,\hat p)\) obtained using Equation~\eqref{eq:final_eq}, we form the component separation weights in the harmonic domain as:
\[
\widehat{\mathbf{W}}
\;\equiv\;
\bigl[\mathbf{\hat A}^{T}\,\mathbf{\widehat N}_{\ell}^{-1}\,\mathbf{\hat A}\bigr]^{-1}
\,\mathbf{\hat A}^{T}\,\mathbf{\widehat N}_{\ell}^{-1}\,.
\]
These weights combine the estimated mixing matrix \(\mathbf{\hat A = A(\hat \beta)}\) and the noise covariance \(\mathbf{\widehat N_\ell = N_\ell (\hat p)}\) to recover the sky maps:
\begin{equation}
\begin{split}
\widehat{\mathbf{s}}_{\ell m}
&=\widehat{\mathbf{W}}\,\mathbf{d}_{\ell m}
=\widehat{\mathbf{W}}\bigl(\mathbf{A}\,\mathbf{s}_{\ell m}+\mathbf{n}_{\ell m}\bigr).\\
\end{split}
\end{equation}
The corresponding angular power spectrum becomes 
\begin{equation}
\label{eq:recovered_aps}
\begin{split}
\mathbf{\widehat{C}_{\ell}^{\rm CMB}}
&=\Bigl\langle\widehat{\mathbf{s}}_{\ell m}\,\widehat{\mathbf{s}}_{\ell m}^{\dagger}\Bigr\rangle
=\Bigl\langle\widehat{\mathbf{W}}\mathbf{d}_{\ell m}\,\mathbf{d}_{\ell m}^{\dagger} \,\widehat{\mathbf{W}}^{T}\Bigr\rangle \\
&=\mathbf{C}_{\ell}^{\rm CMB}+ \mathbf{C}_{\ell}^{\rm stat} +\widehat{\mathbf{W}}\,\mathbf{N}_{\ell}\,\widehat{\mathbf{W}}^{T}.
\end{split}
\end{equation}
The first term corresponds to the input CMB spectrum, the second term corresponds to the statistical foreground residuals due to any mismatch in the spectral parameters recovery due to the presence of residuals of noise~\cite{Errard_2019, Rizzieri_2025b}, and the third term quantifies the residual noise after component separation.  In practice, one estimates and subtracts the residual noise bias \(\widehat{\mathbf{W}}\mathbf{N}_{\ell}\widehat{\mathbf{W}}^{T}\) to obtain an unbiased estimate of \(\mathbf{C}_{\ell}^{\rm CMB}\) in the limit of no foreground residuals \(\mathbf{C}_{\ell}^{\rm stat}\).  These reconstructed spectra can then be fed directly into the forecast for the tensor-to-scalar ratio \(r\).

\subsection{Likelihood on the tensor-to-scalar ratio}  
\label{sec:r_likelihood}

With the reconstructed CMB spectrum \(\mathbf{\widehat{C}_{\ell}^{\rm CMB}}\) in hand, we build the likelihood for the tensor-to-scalar ratio \(r\) following the procedure outlined by LiteBIRD~\cite{2023PTEP.2023d2F01L}.  At each multipole \(\ell\), the B-mode spectrum is modeled as
\begin{equation}
\label{eq:total_Cl}
    \mathbf{
C_{\ell}}(r)=\mathbf{C_{\ell}^{\rm lens}}+r\,\mathbf{C_{\ell}}^{r=1}+\mathbf{C}_{\ell}^{\rm stat} +\mathbf{\widehat N_{\ell}}\,,
\end{equation}
where \(\mathbf{C_{\ell}^{\rm lens}}\) is the lensed-\(\Lambda\)CDM contribution, \(\mathbf{C_{\ell}}^{r=1}\) is the primordial B-mode template for \(r=1\), \(\mathbf{C}_{\ell}^{\rm stat}\) is the statistical foreground residuals, and \(\mathbf{\widehat N_{\ell}}\) is the noise bias from Equation~\eqref{eq:recovered_aps}.

Assuming that \(\mathbf{\widehat{C}_{\ell}^{\rm CMB}}\) follows a scaled-\(\chi^{2}\) distribution with \(N\equiv (2\ell+1) f_{\rm sky}\) degrees of freedom for an observed sky fraction \(f_{\rm sky}\), the per-\(\ell\) likelihood is given by
\begin{equation}
\label{eq:L_r}
\begin{split}
\mathcal{L}_{\ell}(r)
&=\mathcal{P}\bigl(\mathbf{\widehat{C}_{\ell}^{\rm CMB}\mid C_{\ell}}(r)\bigr)\\
&=\frac{1}{2^{N/2}\,\Gamma(N/2)}\,
\Bigl(\tfrac{\displaystyle N}{\displaystyle \mathbf{C_{\ell}}(r)}\Bigr)^{N/2}\,\\
& \times \quad \Bigl( \mathbf{\widehat{C}_{\ell}^{\rm CMB}} \Bigr)^{\,N/2-1}
\exp\,\Bigl[-\tfrac{\displaystyle N\,\mathbf{\widehat{C}_{\ell}^{\rm CMB}}}{\displaystyle 2\, \mathbf{C_{\ell}}(r)}\Bigr].
\end{split}
\end{equation}
The full likelihood over all multipoles is the product \(\mathcal{L}(r)=\prod_{\ell}\mathcal{L}_{\ell}(r)\), from which we compute the upper limits on \(r\). 

Since the theoretical CMB spectrum \(\mathbf{C_{\ell}}(r)\) depends only on \(r\) for a fixed cosmology, we write the data likelihood as \(\mathcal{P}(\mathbf{\widehat{C}_{\ell}^{\rm CMB}}\mid r)\), the posterior distribution can be expressed as
\begin{equation}
\label{eq:P_r}
\mathcal{P}(r)\;\propto\;\prod_{\ell=\ell_{\min}}^{\ell_{\max}}
\mathcal{P}\bigl(\mathbf{\widehat{C}_{\ell}^{\rm CMB}}\mid r\bigr)\,.
\end{equation}
In our work, we consider a uniform prior on \(r\), taking a constant value in the range [-1, 1].
Evaluating this posterior product yields an unbiased estimate of \(r\) as long as the model for recovered CMB spectrum is ideal, which is true in our work, thanks to the fact that we use ensemble-averaged estimators with simplistic foreground simulations and assumptions, limited only by the resolution of the grid or sampler used.  Changes in the noise model directly affect the width of the posterior distribution \(\mathcal{P}(r)\), so this framework provides a clear forecast of how different noise assumptions influence the achievable upper limit on the tensor-to-scalar ratio, in the limit that the spectral parameters are unbiased. In principle, a mismatch in the estimation of the spectral parameters adds bias to this estimate.  We quote the \(95\%\) upper limit of the likelihood on \(r\) as the figure-of-merit, \(r_{95}\), in this work. 

To verify that the upper limits derived from the likelihood are reasonable, we cross-check them using the uncertainty estimated from the Fisher matrix~\cite{fisher1935}. Since there is only one free parameter (\(r\)) considered in our work, the Fisher matrix reduces to a single scalar, given by
\begin{equation}
\label{eq:fisher}
    F_{rr} = \sum_{\ell = \ell_{\min}}^{\ell_{\max}} \frac{(2\ell + 1)\,f_{\rm sky}\left( \mathbf{C}_{\ell}^{r=1} \right)^2 }{2\left(\,\mathbf{C}_{\ell}(r)\right)^2} 
\end{equation}
where \(\mathbf{C}_{\ell}(r)\) is the total CMB power spectrum given in Equation~\eqref{eq:total_Cl}. Applying the Cram\'er-Rao bound, the minimum possible standard deviation on \(r\) is
\begin{equation}
\label{eq:sigma_f}
    \sigma_F(r) \equiv \sqrt{F^{-1}_{rr}}\,.
\end{equation}
The uncertainty estimated from the Fisher information matrix is statistically comparable with the \(1-\sigma\) region derived from the likelihood in Equation~\eqref{eq:L_r}, acting as a cross-check in our study.

\section{Results} \label{sec:results}
We apply our ensemble-average, bias-corrected ridge likelihood to a range of correlated-noise models to understand how the noise parameters \(\alpha\) (the spectral slope) and \(\ell_{0}\) (the knee multipole) influence component separation and, ultimately, the recovered B-mode signal.  All forecasts assume \(\ell_{\max}=256\) and the nominal ECHO instrument specifications (see Table 1 of \cite{Adak_2022}).  We begin by examining the effect of varying \(\alpha\), then explore changes in \(\ell_{0}\), and finally assess their combined impact, together with beam smoothing, on the residual power spectra. We emphasize at this stage that the models are chosen primarily to illustrate the performance of the methodology discussed in this work and are not indicative of the observational limitations of ECHO or other missions. When selecting our models, we opt to build complex models, considering that higher-frequency channels tend to have larger levels of noise, as observed in ground-based experiments such as SPT-3G \cite{2021PhRvD.104b2003D, 2025arXiv250502827Z}. 
Although such pessimistic low-\(\ell\) noise may not 
affect a space-based mission like ECHO, we choose this configuration as an illustration, a proof of concept for our proposed methodology.

\subsection{Effect of noise parameters}

Because our goal is to constrain primordial B-modes, we focus on multipoles \(\ell\le256\), where the largest-scale signal (e.g.\ the reionization bump) resides and smaller scales contribute marginally to \(r\), depending on the level of delensing.  Below, we summarize how each noise parameter alters the shape and amplitude of the residual power spectrum.

\begin{figure}
  \includegraphics[width=0.4\textwidth]{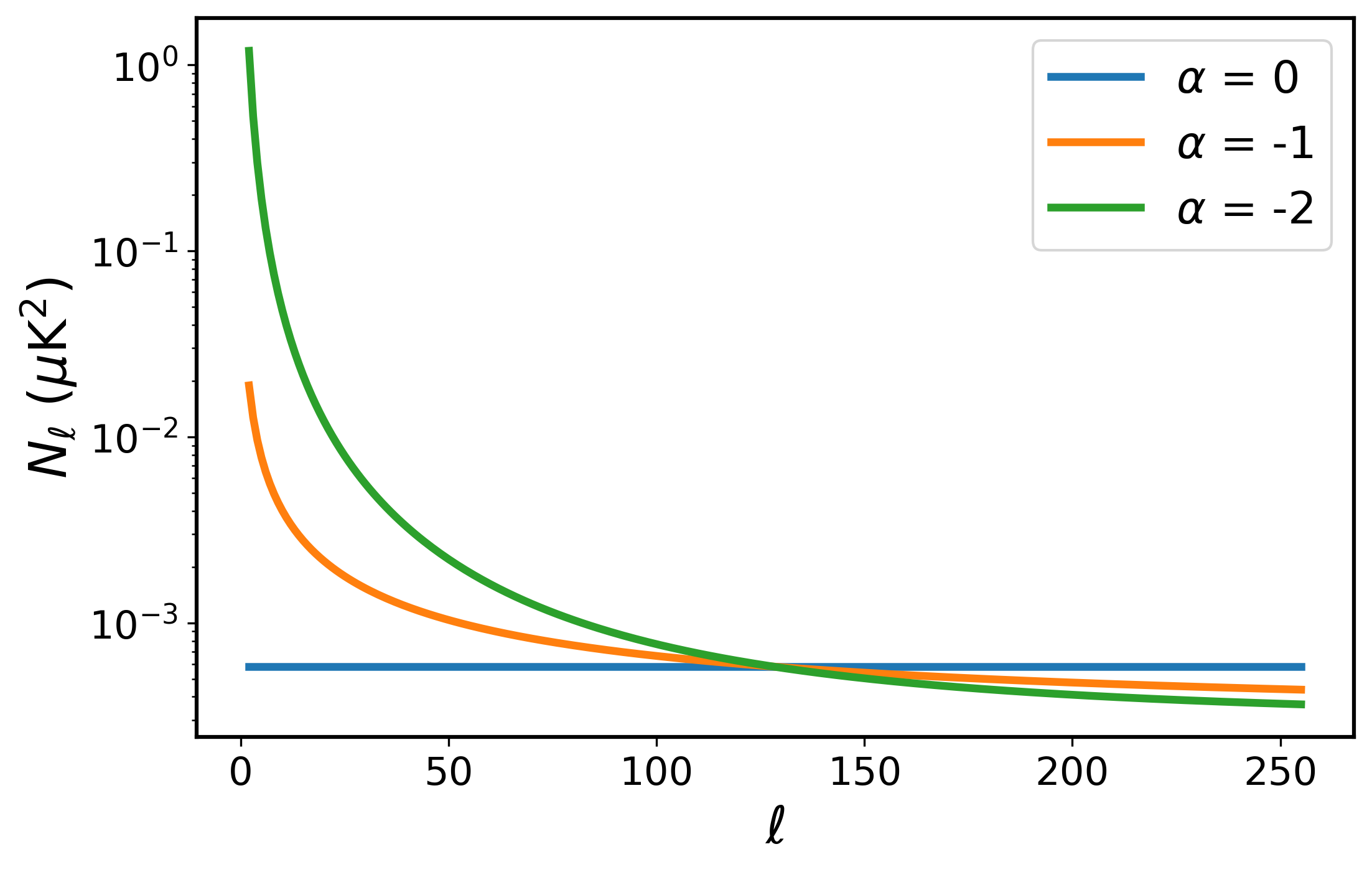}
  \includegraphics[width=0.4\textwidth]{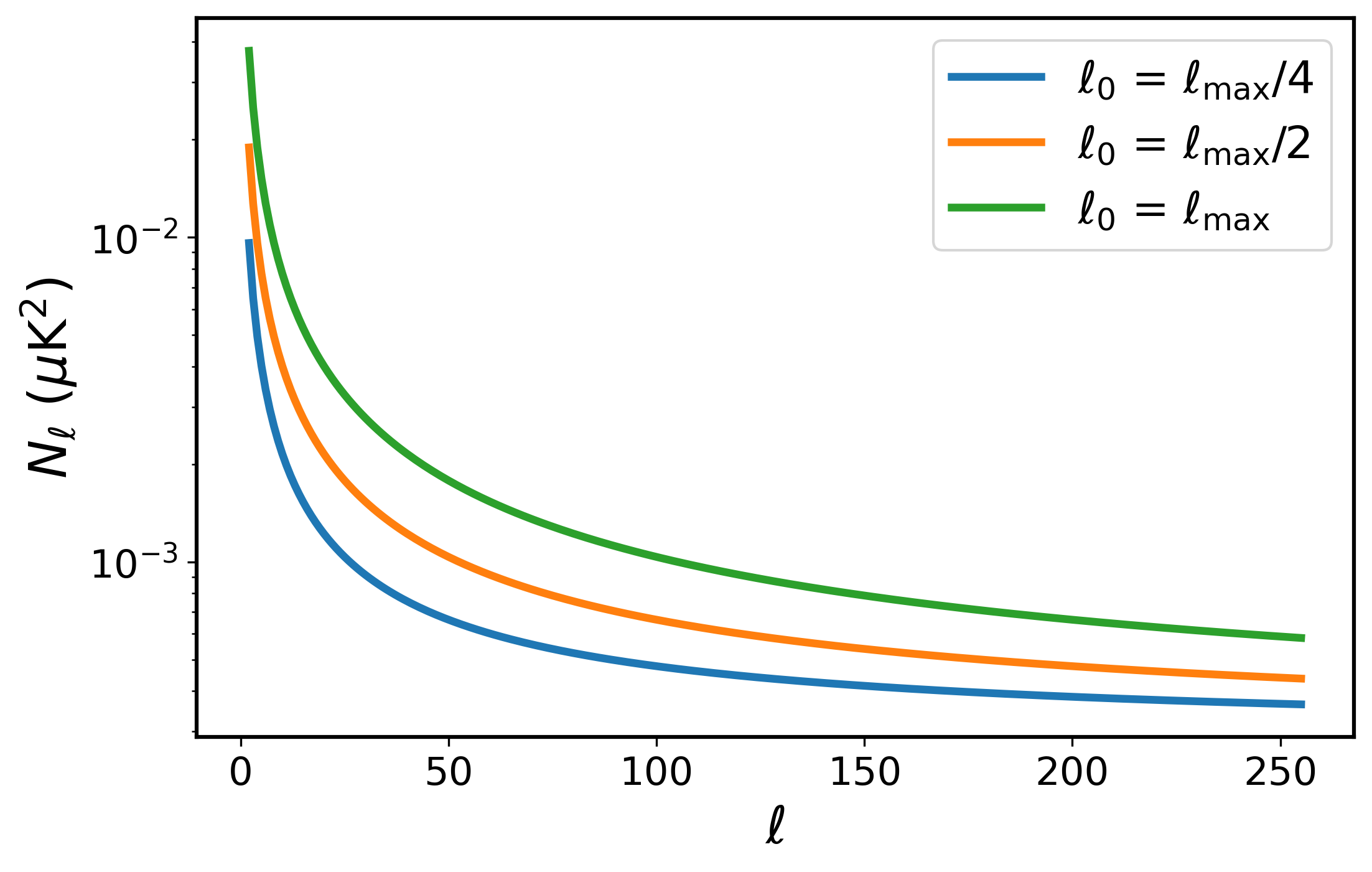}
  \caption{Variation of the noise angular power spectra with changes in the noise parameters \(p\) for fixed \(\sigma_{\rm white}^{2}=1\). \textit{Top:} Changes observed in noise power when the slope \(\alpha\) is varied at constant \(\ell_{0}=128\). Steeper (more negative) \(\alpha\) values boost power at low multipoles, then converge to the white-noise level for \(\ell>\ell_{0}\). \textit{Bottom:} Changes observed for different knee scales \(\ell_{0}\), with fixed \(\alpha=-1\).  Increasing \(\ell_{0}\) extends the correlated-noise regime to higher multipoles, reducing the signal-to-noise ratio over those scales.}
  \label{fig:noise_aps_p}
\end{figure}

\subsubsection{Power-law index $\alpha$}

The parameter \(\alpha\) controls the low-frequency noise behavior, ranging from white (\(\alpha=0\)) through pink (\(\alpha=-1\)) to red noise (\(\alpha= -2\) or smaller).  As \(\alpha\) becomes more negative, the spectrum steepens at low \(\ell\) and can far exceed the white noise floor; see top panel of Figure~\ref{fig:noise_aps_p}.  Beyond the knee multipole, \(\ell\gg\ell_{0}\), the power returns to the white-noise level.  In the red-noise regime (\(\alpha\lesssim-2\)), low-\(\ell\) contamination can dominate even over diffuse foregrounds, making it challenging to recover features such as the reionization bump in the B-mode spectrum.

In this work, four models are considered to study the impact of $\alpha$ on component separation at a constant value of $\ell_0 = \ell_{\max}/2$:

\begin{itemize}
    \item $\alpha = 0$ --- White Noise
    \item $\alpha = -1$ --- Pink Noise
    \item $\alpha = -2$ --- Red Noise
    \item $\alpha = \mathrm{Variable}$ --- Linearly changes across frequency channels between -1 and -5 in steps of 0.2.
\end{itemize}

\begin{figure}
  \centering
  \includegraphics[width=\columnwidth]{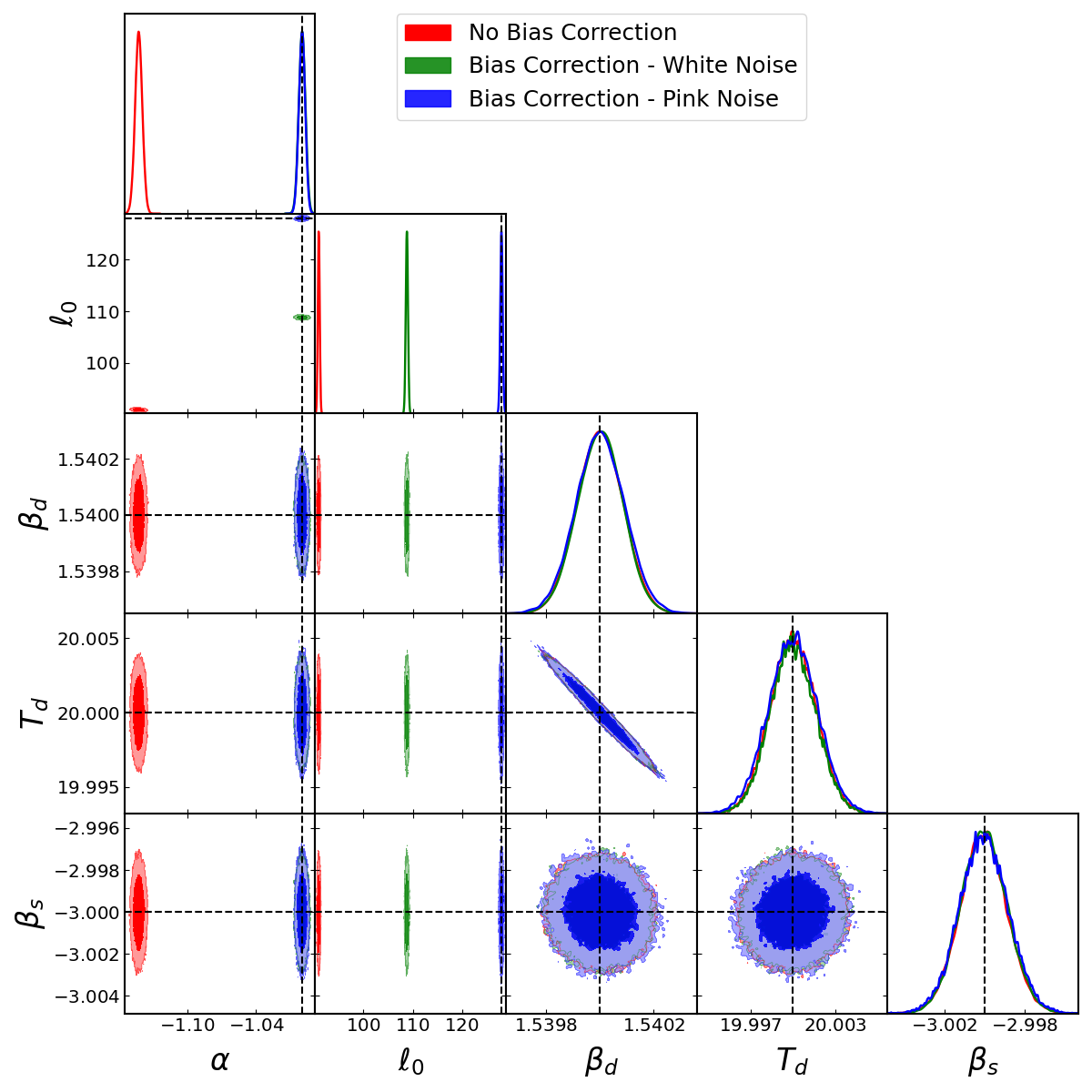}
  \caption{Joint posterior contours (68 \% and 95 \% confidence levels) for the noise parameters \((\alpha,\,\ell_{0})\) and foreground spectral parameters (\(\beta_{d}\), \(T_{d}\), \(\beta_{s}\)) obtained by sampling different likelihoods --
  the ensemble-averaged version of the original ridge likelihood (Equation \eqref{eq:RidgeLikelihood}) without bias correction(\textit{red}), the generalized bias-corrected ridge likelihood (Equation \eqref{eq:final_eq}) assuming white noise for bias correction as per Equation \eqref{white_noise_th} (\textit{green}), and the generalized bias-corrected ridge likelihood (Equation \eqref{eq:final_eq}) assuming exact knowledge of input noise as per Equation \eqref{eq:pink_noise_th} (\textit{blue}).  Dashed lines mark the true input values: \(\alpha=-1\), \(\ell_{0}=128\) for the noise model, and \(\beta_{d}=1.54\), \(T_{d}=20\text{ K}\), \(\beta_{s}=-3\) for the foregrounds. The spectral parameters are recovered identically, while the performance on the noise parameters differs, with a reduction in bias seen when more information about the noise is given.}
  \label{fig:triangle_plot}
\end{figure}

For each noise scenario, we minimize the ensemble-average ridge likelihood (Equation~\ref{eq:final_eq}) to recover both noise and foreground parameters.  As an illustrative case, Figure~\ref{fig:triangle_plot} displays the joint posterior contours for \(\alpha=-1\) and \(\ell_{0}=128\) as inputs, for different choices of bias correction. If we consider that we have access to the true noise as in Equation \eqref{eq:pink_noise_th}, then we see that the noise parameters are recovered without any bias, as depicted by the blue curves in the figure. When we adopt a reasonable assumption to use white noise as the bias correction, as in Equation \eqref{white_noise_th}, we see that the noise parameters, especially the knee-multipole \(\ell_0\), suffer a slight bias, as depicted by the green curves. When we compare this result with those obtained without bias correction, as given by the red curves, we find that a white noise assumption is justified.

Interestingly, the spectral parameters are unaffected by the choice of bias correction for this simple case considered, where the noise parameters are constant across frequency channels. For complicated cases, either as shown in Figure \ref{fig:incorrect_estimate}, or as discussed later, we see biases in the spectral parameters.

\begin{figure}
  \centering
  \includegraphics[width=\columnwidth]{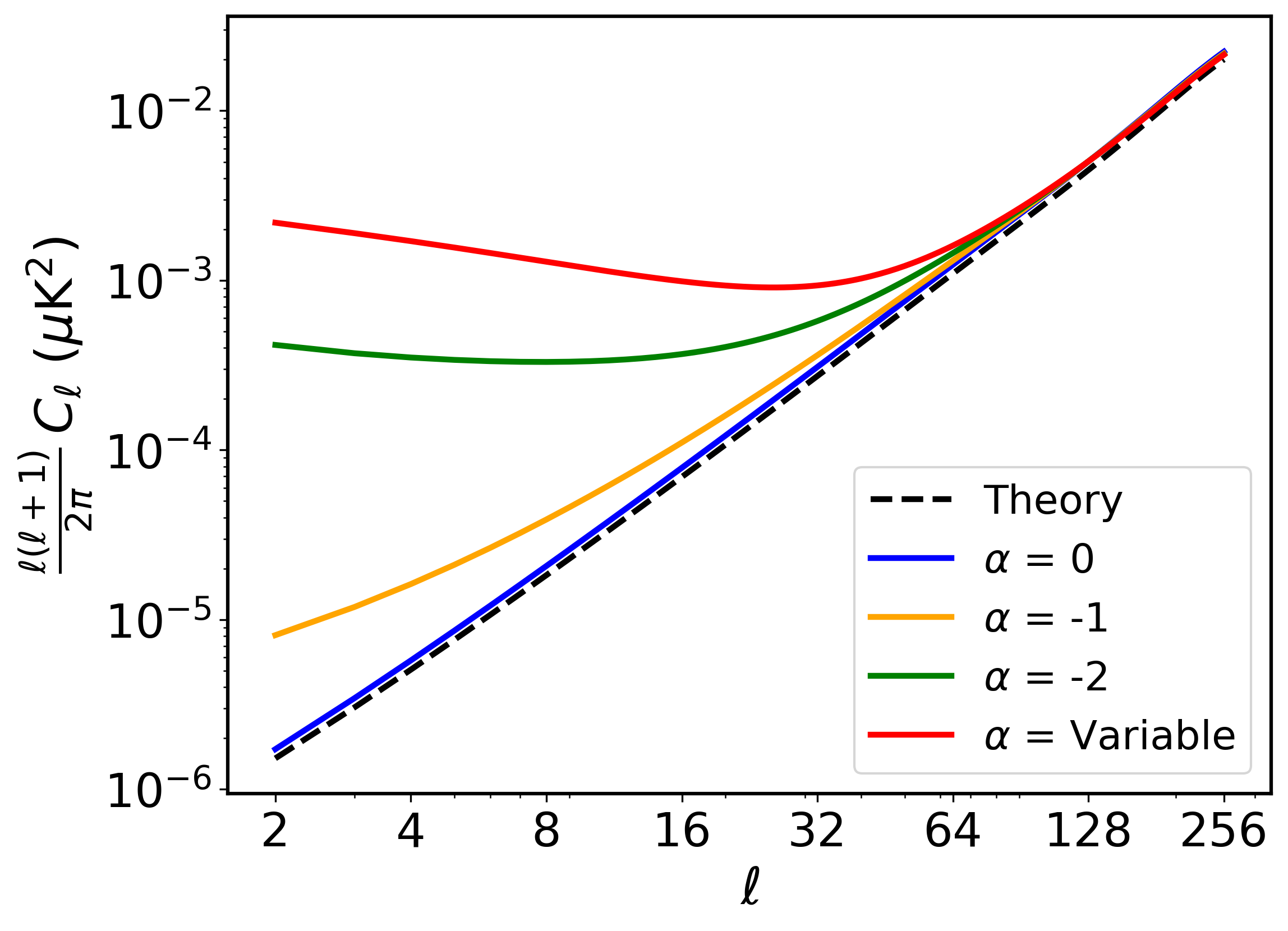}
  \caption{Recovered CMB angular power spectra \(\mathbf{\widehat{C}_{\ell}^{\rm CMB}}\) for each noise slope \(\alpha\) at a constant knee multipole \(\ell_0 = \ell_{\rm max}/2\), where \(\ell_{\rm max} = 256\).  The black dashed line shows the theoretical CMB spectrum, and colored lines correspond to \(\alpha=0\), \(-1\), \(-2\), and a channel-dependent \(\alpha\) spanning \(-1\) to \(-5\).  Larger \(|\alpha|\) values increase residual noise power at low multipoles, causing greater deviations from the theoretical curve.}
  \label{fig:rec_alpha}
\end{figure}

Once we have recovered the noise and spectral parameters, we then use the posterior means to form the weight matrix \(\widehat{\mathbf{W}}\) and reconstruct the CMB maps \(\widehat{\mathbf{s}}_{\ell m}\).  Figure~\ref{fig:rec_alpha} compares the recovered power spectra across the four \(\alpha\) models.  Steeper noise slopes (\(|\alpha|\) larger) clearly elevate the residual power at low multipoles, increasing the deviation of \(\mathbf{\widehat{C}_{\ell}^{\rm CMB}}\) from the theoretical spectrum as noise complexity grows.

\begin{figure*}
  \centering
  \includegraphics[width=0.32\textwidth, height = 5.4cm]{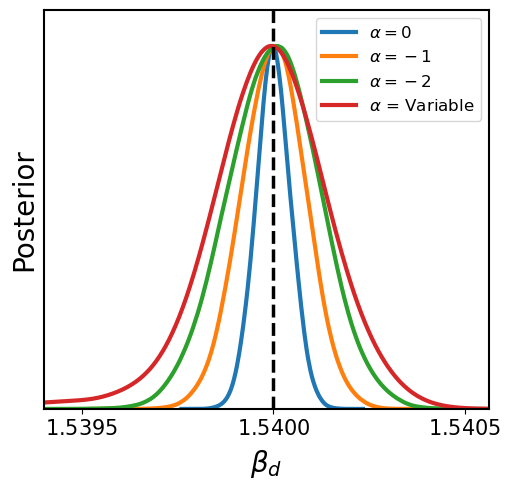}
  \includegraphics[width=0.32\textwidth, height = 5.4cm]{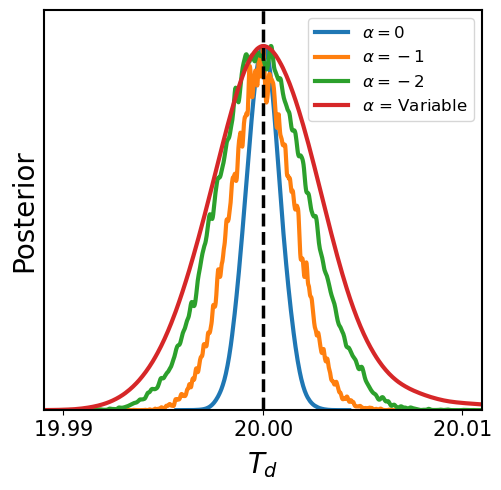}
  \includegraphics[width=0.32\textwidth]{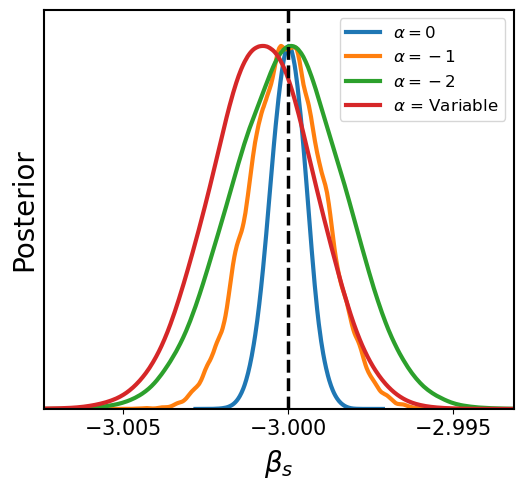}
  \caption{Marginalized posteriors for the dust spectral index \(\beta_{d}\) (left), dust temperature \(T_{d}\) (center), and synchrotron index \(\beta_{s}\) (right) under four noise-slope models: \(\alpha=0\), \(-1\), \(-2\), and a channel-dependent \(\alpha\) spanning \(-1\) to \(-5\).  All assume \(\ell_{0}=128\) and the generalized ridge likelihood (Equation~\ref{eq:final_eq}).  Although unbiased, greater \(|\alpha|\) leads to broader posteriors, indicating increased uncertainty due to stronger low-\(\ell\) noise correlations.}
  \label{fig:fgparams_alpha}
\end{figure*}

Additionally, Figure~\ref{fig:fgparams_alpha} presents the one-dimensional marginalized posteriors for each foreground spectral parameter \(\beta\) under the four \(\alpha\) models.  Although all distributions peak at the true input values, their widths increase systematically with \(|\alpha|\).  This broadening reflects the shrinking signal-to-noise ratio as noise correlations grow, which reduces the information available for accurate parameter recovery and thus raises the posterior uncertainties. We see that the synchrotron slope index \(\beta_s\) is slightly more susceptible to the level of noise as the complicated model of variable \(\alpha\) causes a small shift in its peak, but as this is well within the \(1-\sigma\) region of the other models, this cannot be classified as any significant bias.

\subsubsection{Varying the knee multipole \(\ell_{0}\)}
The knee scale \(\ell_{0}\) sets the transition between correlated low-\(\ell\) noise and the white-noise regime.  For \(\alpha\le0\), all multipoles \(\ell\le\ell_{0}\) experience enhanced noise power due to the \(1/\ell^{\alpha}\) term.  As \(\ell_{0}\) increases, this correlated-noise regime extends to higher multipoles, thereby lowering the signal-to-noise ratio over a broader range of angular scales. The bottom panel of Figure~\ref{fig:noise_aps_p} illustrates how larger values of \(\ell_{0}\) raise the noise floor across \(\ell\le\ell_{0}\), before the spectrum returns to the white-noise level at \(\ell>\ell_{0}\).

We next fix the noise slope at \(\alpha=-1\) and consider different models in which we vary the knee multipole \(\ell_{0}\):
\begin{itemize}
    \item $\ell_0 = \ell_{\max}/4 = 64$
    \item $\ell_0 = \ell_{\max}/2 = 128$
    \item $\ell_0 = \ell_{\max} = 256$
    \item $\ell_0 = \mathrm{Variable}$ --- Linearly changes across frequency channels between \(\ell_{\min} = 2\) and \(\ell_{\max} = 256\), in steps of \(\sim\) 13-14.
\end{itemize}
It is important to note that these values are not physically motivated. For a typical space mission, for instance, such values would generically be way too high. Yet, we consider such cases to test and validate our approach. 
For the complicated model of variable \(\ell_0\), we chose linearly changing values which are then rounded off to the nearest integer, starting from \(\ell_0 = 2\) in the lowest frequency channel to \(\ell_0 =  256\) in the largest frequency channel. The rounding off results in uneven step-sizes, which alternate between 13 and 14 across successive frequency channels.

\begin{figure}
  \centering
  \includegraphics[width=\columnwidth]{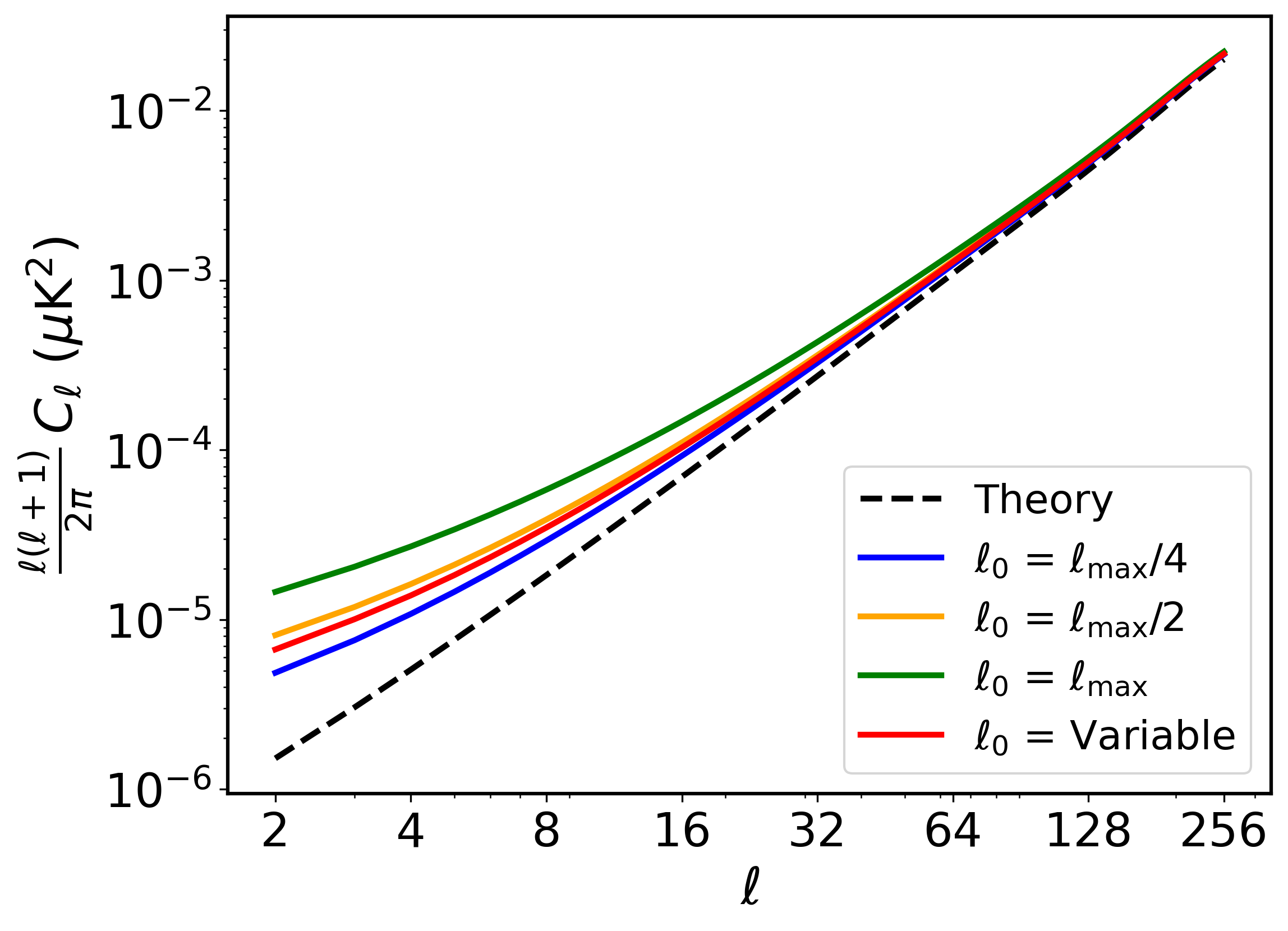}
    \caption{Recovered CMB angular power spectra \(\mathbf{\widehat{C}_{\ell}^{\rm CMB}}\) for each knee multipole \(\ell_0\) at a noise slope \(\alpha=-1\). The black dashed line is the theoretical CMB spectrum; colored lines correspond to \(\ell_{0}=64, 128, 256\)  and a channel-dependent \(\ell_0\) spanning from \(2\) to \(\ell_{\max} = 256\). Larger \(\ell_{0}\) values increase low-\(\ell\) residuals, causing greater departures from the true spectrum.}
  \label{fig:rec_ell0}
\end{figure}

Figure~\ref{fig:rec_ell0} shows that increasing \(\ell_{0}\) extends the correlated-noise regime to higher \(\ell\), which in turn amplifies residual power at \(\ell<20\). For each scenario, we minimize the ensemble-average ridge likelihood (Equation~\ref{eq:final_eq}) to recover best-fit noise and foreground parameters.  The resulting CMB power spectra consistently deviate at low multipoles in proportion to the noise complexity.

In the mixed-\(\ell_{0}\) model, where \(\ell_{0}\) varies linearly across frequency channels, some frequency bands adopt very low knee multipoles, causing the noise spectrum to return quickly to the white-noise floor above those scales.  This leads to lower overall residual power and smaller parameter uncertainties than in the constant \(\ell_{0}\) cases, where every channel shares a high knee scale and sustains elevated noise over a broader range of \(\ell\).  Even so, the impact of varying \(\ell_{0}\) remains modest compared to changes in the noise slope \(\alpha\).  Because \(\alpha\) enters as an exponent in the power-law, small shifts in its value multiply the noise power more significantly at a given angular scale, producing larger deviations from the true CMB signal.

\begin{figure*}
  \centering
  \includegraphics[width=0.32\textwidth]{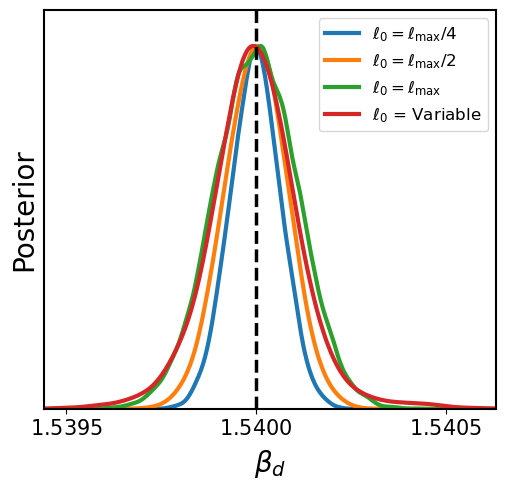}
  \includegraphics[width=0.32\textwidth, height = 5.5cm]{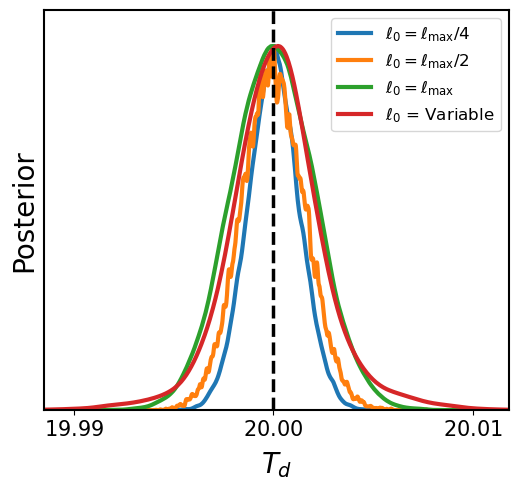}
  \includegraphics[width=0.32\textwidth]{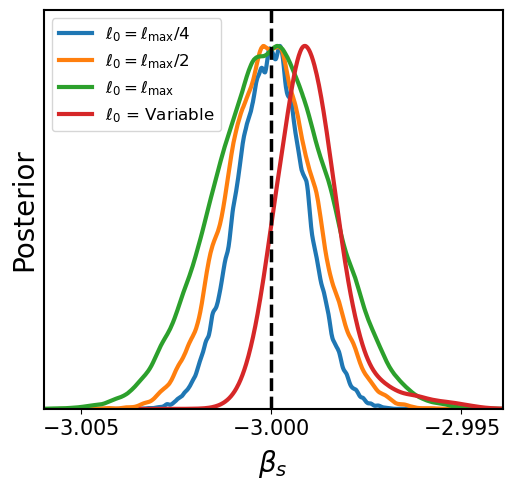}
  \caption{Marginalized posteriors for \(\beta_{d}\) (left), \(T_{d}\) (center), and \(\beta_{s}\) (right) under four \(\ell_{0}\) models with \(\alpha=-1\).  As \(\ell_{0}\) increases, the distributions widen due to stronger low-\(\ell\) noise correlations, indicating increased parameter uncertainty.}
  \label{fig:fgparams_ell0}
\end{figure*}
  
Likewise, the marginalized posteriors for the spectral parameters \(\beta\) broaden with larger \(\ell_{0}\), reflecting the reduced signal-to-noise ratio (Figure~\ref{fig:fgparams_ell0}), yet remain centered on the true inputs. Similar to the case with the variation in \(\alpha\), the more complicated model of variable \(\ell_0\) produces a small shift in the distribution of the synchrotron slope index \(\beta_s\), but well within the \(1-\sigma\) region of the other models considered.

\subsubsection{Varying $\alpha$, $\ell_0$ and including beam convolution}

Thus far, we have examined noise models with either a fixed slope (\(\alpha\)) or a fixed knee multipole (\(\ell_{0}\)) across all frequency channels.  To capture a more realistic scenario, we combine these approaches into a single, generalized model in which both \(\alpha\) and \(\ell_{0}\) vary linearly with frequency. This linearly varying model is constructed by combining the previous two models of variable \(\alpha\) and variable \(\ell_0\). This model is used both to generate the mock data and to recover the parameters, allowing each channel to exhibit its own correlated-noise behavior.

In practice, the measured sky maps are also smoothed by the instrument beam, which we approximate as Gaussian with a known Full-Width at Half-Maximum (FWHM).  Convolution by the beam suppresses power at scales smaller than the FWHM, and imperfect deconvolution of its Gaussian tail can introduce excess power at high multipoles.  This beam-induced bias, when combined with the channel-dependent noise correlations, yields our most complex noise model~\cite{Rizzieri_2025}. 

\begin{figure}
  \centering
  \includegraphics[width=\columnwidth]{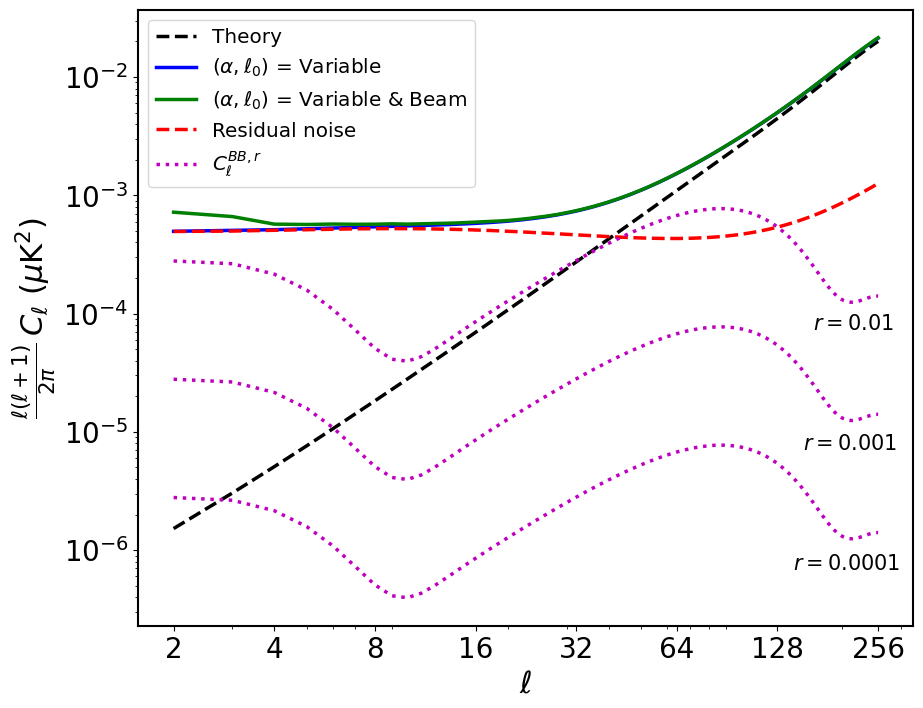}
  \caption{Recovered CMB angular power spectra \(\mathbf{\widehat{C}_{\ell}^{\rm CMB}}\) for the generalized noise model, with beam smoothing to 39.9\(' \) FWHM (green solid) and without (blue solid).  The residual-noise spectrum (red dashed) dominates the primordial signal (black dashed) at low multipoles. Dotted magenta lines show the primordial B-mode power spectrum for different values of \(r\ \in \{0.01, 0.001, 0.0001\}\).}
  \label{fig:rec_gm}
\end{figure}

In our analysis, each frequency channel is first convolved with its native Gaussian beam and then smoothed to a common resolution of 39.9\('\) FWHM, corresponding to the largest ECHO beam specification.  Figure~\ref{fig:rec_gm} compares the recovered CMB power spectra for the generalized noise model with (green) and without (blue) this final beam smoothing.  The two curves overlap almost perfectly across \(\ell\le 256\), differing up to \(\sim 1\%\) at the very lowest multipoles and becoming negligible at higher \(\ell\). As expected, there is a slight excess of power at smaller scales arising from the Gaussian tail of the beam.  

Meanwhile, the residual-noise spectrum (red dashed line) exceeds the theoretical CMB B-mode signal (black dashed) at low \(\ell\), demonstrating that correlated noise alone can overwhelm the primordial signal even when foreground residuals are absent.  This case thus highlights the critical importance of accurately controlling both noise correlations and beam effects in any component separation strategy. The primordial B-mode power spectrum \(r\,\mathbf{C_{\ell}}^{r=1}\) for different values of \(r\) is also shown to highlight that the noise bias makes it difficult to retrieve the cosmological signal perfectly. But the inclusion and modeling of the noise bias \(\mathbf{\widehat N_{\ell}}\) in the likelihood ensures that this does not contribute to a huge bias in the forecasts made on the figure-of-merit,  \(r_{95}\). 

\begin{figure*}
  \centering
  \includegraphics[width=0.32\textwidth]{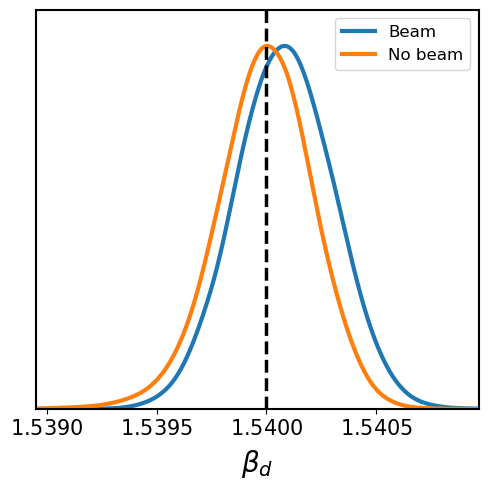}
  \includegraphics[width=0.32\textwidth]{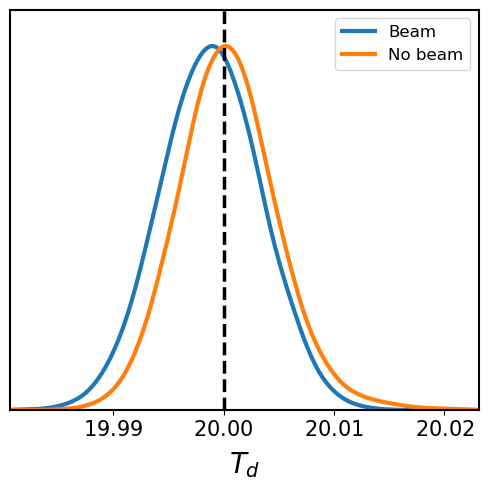}
  \includegraphics[width=0.32\textwidth]{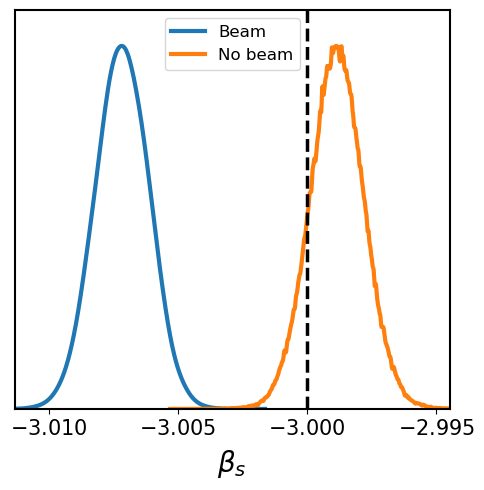}
  \caption{Marginalized posterior distributions for the dust index \(\beta_{d}\) (left), dust temperature \(T_{d}\) (center), and synchrotron index \(\beta_{s}\) (right) under the generalized noise model, comparing results with beam smoothing (blue) to those without (orange), along with the true input values (black dashed). Beam effects produce only minimal changes in the recovered parameter uncertainties.}
  \label{fig:fgparams_gm}
\end{figure*}

The foreground spectral parameters \(\beta\) are again recovered without any significant bias, albeit with slightly broader posterior distributions under the generalized noise model (see Figure~\ref{fig:fgparams_gm}).
We observe that, for all the spectral parameters, there is a slight mismatch between the two cases considered, which differ by the inclusion of the beam. While the distributions of the spectral parameters for thermal dust (\(\beta_d, T_d\)) have considerable overlap between the two cases, the synchrotron slope index \(\beta_s\) shows no overlap for the posteriors. Since the distributions are intrinsically very narrow, this does not result in a large bias in its value. These changes observed in the distributions can be attributed to the convolution/deconvolution effects of the beam, as well as the fact that the noise model chosen consists of frequency-dependent parameters. 
The comparatively larger bias observed in $\beta_s$ relative to the other spectral parameters may stem from the fact that the low-frequency beams sizes are closer to the common beam, potentially leading to imperfections in the common-beam convolution. In principle, such a bias could be mitigated by explicitly incorporating the beam into the component separation formalism~\cite{Rizzieri_2025}.

In contrast, if we were to assume the noise covariance to be white (consider Equation~\eqref{eq:original_likelihood} for component separation), then the recovery of the spectral parameters do not converge, even when we extend the priors to unphysical limits like \( | \hat \beta - \beta| \gg 10^2\) or \( \beta_s > 0\). This highlights a significant limitation of the traditional methodology when dealing with non-trivial noise configurations.

In summary, variations in the noise slope \(\alpha\) have the most pronounced effect on low-\(\ell\) residual power, while shifts in the knee multipole \(\ell_{0}\) yield smaller but still measurable changes.  Adding both noise parameters together, and including realistic beam smoothing, further degrades performance at the largest scales; nonetheless, our bias-corrected ridge-likelihood framework captures these trends accurately and efficiently, thanks to the fact that our model, the parametrization of $\mathbf{N}_\ell$, has the correct form.

\subsection{Estimates of the tensor-to-scalar ratio}

Having characterized how correlated noise and beam effects impact the recovered B-mode spectra, we turn to their influence on constraining the tensor-to-scalar ratio \(r\).  Using the per-\(\ell\) likelihood of Equation~\eqref{eq:L_r} and the posterior product of Equation~\eqref{eq:P_r}, we compute the \(95\%\) upper limit \(r_{95}\) for each noise model, taking the theoretical CMB spectrum as the unbiased reference. In order to complete a statistical cross-check, we use Equations~\ref{eq:fisher} and \ref{eq:sigma_f} to compute the uncertainty provided by the Fisher matrix.

\begin{table}[!htbp]
\centering
\begin{tabular}{|c|c|c|c|c|}
\hline
\(\alpha\) & \(\ell_{0}\) & Beam & \(r_{95} \; (f_{\rm sky} = 1)\) & \(r_{95} \; (f_{\rm sky} = 0.5)\) \\
\hline
0     & 128  & No  & \(7.8\times10^{-5}\)  & \(1.37\times10^{-4}\)   \\
-1    & 64  & No  & \(1.55\times10^{-4}\) & \(2.47\times10^{-4}\)  \\
-1    & Variable & No  & \(1.93\times10^{-4}\) & \(3.01\times10^{-4}\)\\
-1    & 128  & No  & \(2.12\times10^{-4}\)  & \(3.27\times10^{-4}\) \\
-1    & 256  & No  & \(2.99\times10^{-4}\)  & \(4.45\times10^{-4}\)\\
-2    & 128  & No  & \(4.55\times10^{-4}\)  & \(6.49\times10^{-4}\)\\
Variable & Variable & No  & \(4.95\times10^{-4}\) & \(7.02\times10^{-4}\) \\
Variable & Variable & Yes & \(5.22\times10^{-4}\) & \(7.32\times10^{-4}\)  \\
Variable & 128  & No  & \(5.25\times10^{-4}\) & \(7.47\times10^{-4}\)\\
\hline
\end{tabular}
\caption{Projected 95\% upper limits \(r_{95}\) for various noise model configurations obtained by varying the noise parameters \(p =  (\alpha, \, \ell_0)\), the beam and the fraction of sky (\(f_{\rm sky}\)) observed. ``Variable'' indicates channel-dependent values of \(\alpha\) or \(\ell_{0}\). When only half the sky is viewed (\(f_{\rm sky} = 0.5\)), estimated values of \(r_{95}\) are higher by \(\sim 1/\sqrt{f_{\rm sky}} = \sqrt{2}\).}
\label{tab:r95}
\end{table}

Table~\ref{tab:r95} shows that moving from the simplest (white noise, \(\alpha=0\), \(\ell_{0}=128\)) to the most complex model (both \(\alpha\) and \(\ell_{0}\) varying across channels, plus beam smoothing) degrades the \(r_{95}\) limit by roughly an order of magnitude, even for a full sky (\(f_{\rm sky} = 1\)).  This demonstrates that, even in the absence of significant foreground residuals, unmodeled noise correlations can dominate the uncertainty on \(r\).  

Considering the effect of $\alpha$ on $r_{95}$, it is clear that as the steepness of the power law increases, there is a significant increase in the width of the posterior distribution of $r$, as shown in Figure \ref{fig:fgparams_alpha}. As noise transitions from white to pink to red, there is at least a doubling of  $r_{95}$ at each stage. The knee multipole $\ell_0$ has a less severe impact on the $r_{95}$. For a fixed value of $\alpha$, increasing $\ell_0$ does increase the $r_{95}$, but this change is only about $20 - 30\%$ between models, as shown in Figure \ref{fig:fgparams_ell0}. 

Another takeaway is that the largest value of $r_{95}$ comes from the model in which \(\alpha\) is variable at a constant \(\ell_0 = 128\). This model produces larger residuals compared to the generalized models, including the one with the beam, as those models contain a large fraction of frequency channels with \(\ell_0 < 128\), resulting in a reduction of the overall noise contributed by that channel. This can also be understood by comparing the angular power spectrum of the generalized model (Figure \ref{fig:rec_gm}), which has lesser deviation at the large angular scales, with the model in which \(\alpha\) is variable (red curve in Figure \ref{fig:rec_alpha}).

Choice of complex foreground models would lead to higher residuals, necessitating typically either masking regions close to the galactic plane and consequently decreasing the fraction of sky observed, or using more foreground parameters as with multi-resolution component separation~\cite{2023PTEP.2023d2F01L}. Introducing additional foreground parameters generally leads to increased statistical uncertainties, which in turn result in higher levels of residual foreground contamination. This degrades the final constraint on the tensor-to-scalar ratio \(r\). The fraction of sky observed will manifest as an additional coefficient in the likelihood, as shown in \cite{2023PTEP.2023d2F01L}, and will increase the estimate obtained on the uncertainty in the measurement of \(r\). In the last column of Table~\ref{tab:r95}, we try to understand this effect of masking in order to get a sense of how decreasing the fraction of sky observed (\(f_{\rm sky}\)) will influence the estimation of \(r_{95}\). 

To construct an approximate picture, we choose \(f_{\rm sky} = 0.5\) in this representative example, a typical choice in contemporary CMB experiments \cite{2023PTEP.2023d2F01L}, where heavily contaminated regions such as the Galactic plane are excluded. The results from Table \ref{tab:r95} clearly show that reducing the sky fraction leads to an increase in \(r_{95}\) approximately by a factor of \(1/ \sqrt{f_{\rm sky}}\), driving the upper limits of the likelihood close to the \(10^{-3}\) range, underscoring the trade-off between sky coverage and statistical sensitivity.

Another interesting takeaway is the low values of $r_{95}$, which are as small as $10^{-5} - 10^{-4}$. This can be attributed to, in addition to large $f_{\rm sky}$, the absence of foreground residuals due to the simple choice of foreground model \texttt{d0s0}. Foreground residuals are also low because the spectral parameters are recovered accurately due to our choice of a reasonably good noise model in the component separation. Without this, there would be significant biases on the spectral parameters, leading to systematic residuals which translate towards large biases on \(r\). Since the spectral parameters are accurately estimated for each model (as shown in Figure \ref{fig:triangle_plot}), there is no need for masking, which would not be feasible under realistic observing conditions.

In order to verify that such low values on the upper limits of the likelihood are plausible, we perform a cross-check in Table \ref{tab:fisher_vs_L} by comparing the \(68\%\) CL obtained from the likelihood \(\mathcal{L}_\ell(r)\) (Equation \eqref{eq:L_r}) with the \(1\sigma\) uncertainty obtained from the Fisher matrix, Equation~\eqref{eq:sigma_f}. We note that both $68\%$ CL of  $\mathcal{L}_\ell(r)$ and $\sigma_F(r)$ are in agreement here.\\

\begin{table}[!htbp]
\centering
\begin{tabular}{|c|c|c|c|c|}
\hline
\(\alpha\) & \(\ell_{0}\) & Beam & \(\sigma_F(r)\) & \(68\%\) CL of  \(\mathcal{L}_\ell(r)\)\\
\hline
0     & 128  & No  & \(3.4\times10^{-5}\) &\(3.3\times10^{-5}\)  \\
-1    & 64  & No  & \(7.2\times10^{-5 }\) &\(7.8\times10^{-5}\)   \\
-1    & Variable & No  & \(9.17\times10^{-5}\) &\(1.03\times10^{-4}\)\\
-1    & 128  & No  & \(1.01\times10^{-4}\) &\(1.17\times10^{-4}\)  \\
-1    & 256  & No  & \(1.46\times10^{-4}\) &\(1.82\times10^{-4}\)\\
-2    & 128  & No  & \(3.2\times10^{-4}\) &\(3.23\times10^{-4}\)  \\
Variable & 128  & No  & \(2.65\times10^{-4}\) &\(3.73\times10^{-4}\) \\
Variable & Variable & No  & \(2.49\times10^{-4}\) &\(3.51\times10^{-4}\)\\
Variable & Variable & Yes & \(2.77\times10^{-4}\) &\(3.49\times10^{-4}\)  \\
\hline
\end{tabular}
\caption{Comparison of the \(1 \sigma\) uncertainties provided by the Fisher matrix \(\sigma_F(r)\) (Equation \eqref{eq:sigma_f}) with the  \(68\%\) CL of  \(\mathcal{L}_\ell(r)\) (Equation \eqref{eq:L_r})) for various noise model configurations obtained by varying the noise parameters \(p =  (\alpha, \, \ell_0)\) and the beam, viewed at full sky (\(f_{\rm sky} = 1\)). The uncertainties are quite small due to various considerations -- a simple foreground mode \texttt{d0s0}, a reasonable choice of noise model for estimation, and accurate recovery of spectral parameters.} 
\label{tab:fisher_vs_L}
\end{table}

\section{Conclusions}  
\label{sec:conclusions}

To deal with biases arising from the possible mismodeling of the noise covariance in typical parametric component separation methods, we have proposed an extension of the parametric, maximum-likelihood component separation framework to include revised modeling of the correlated \(1/f^\alpha\) noise in harmonic space.  By introducing a power-law noise covariance and embedding it in a modified ridge likelihood with analytic bias correction, we overcome the limitations of the purely white noise spectral likelihood and ensure unbiased recovery of both foreground and noise parameters. We discuss limitations in assuming models for the theoretical noise power spectrum for bias correction and explore practical avenues that can circumvent this issue and still give reasonable results.

Our ensemble-average pipeline replaces expensive Monte Carlo simulations with the use of traces over theoretical covariances, preserving statistical rigor while reducing computational cost. We illustrate our implementation and validation framework on the ECHO mission.  This approach forecasts the 95\% upper limit of the tensor-to-scalar ratio, \(r_{95}\), under a variety of noise scenarios.  Even in the absence of foreground residuals, we find that correlated noise alone can degrade the \(r_{95}\) upper limit by an order of magnitude, from \(10^{-5}\) to \(10^{-4}\), highlighting the need for a strong control and an accurate modeling of the noise.

The results validate ECHO's baseline noise requirements, demonstrating that its multi-frequency, large-format arrays should achieve \(r_{95}\,\lesssim\,10^{-3}\), even without performing any delensing.  Sensitivity gains of order 10 \% could be realized by modest adjustments to channel bandwidths or beam widths, providing concrete guidance for the mission's instrument design.

Looking ahead, the next step is to extend this framework to more realistic sky and noise models, including spatially varying foregrounds and partial sky coverage, as discussed in \cite{Rizzieri_2025b, Kabalan_2025}, which will introduce pixel-space correlations and masking effects. The noise model considered for bias correction is motivated by the technological limitations at hand. Replacing this model for a more practical or flexible one can lead to better constraints on the noise and spectral parameters and mitigate any further biases. Incorporating delensing techniques or hybrid blind-parametric methods could further tighten constraints on \(r\) and test the robustness of our bias correction in the presence of complex systematics. 

In summary, our noise-aware, bias-corrected methodology equips the new-generation CMB polarization experiments with the analysis tools needed to extract the faint primordial B-mode signal, bringing us closer to uncovering the physics of cosmic inflation.  

\begin{acknowledgements}
    The authors thank the entire SCIPOL team and Radek Stompor in particular for useful discussions and acknowledge funding from the SCIPOL project \footnote{\url{scipol.in2p3.fr}} funded by the European Research Council (ERC) under the European Union's Horizon 2020 research and innovation program (PI: Josquin Errard, Grant agreement No.~101044073). Some of the computations were performed using HPC resources provided by the \textit{Jean Zay} supercomputer at IDRIS under allocation 2024-AD010414161R2 granted by GENCI.
\end{acknowledgements}

\def\bibsection{\section*{References}}
\bibliographystyle{apsrev}
\bibliography{apssamp}

\end{document}